\documentclass[preprint,showpacs,preprintnumbers,amsmath,amssymb]{revtex4}

\usepackage{graphicx}
\usepackage{dcolumn}
\usepackage{bm}

\begin{document}


\title{Leading Order Calculation of Electric Conductivity \\ in Hot Quantum Electrodynamics from Diagrammatic Methods}


\author{Jean-S\'{e}bastien Gagnon}
\email{gagnonjs@physics.mcgill.ca}

\author{Sangyong Jeon}%
\email{jeon@physics.mcgill.ca}

\affiliation{Physics Department, McGill University, 3600 University street, Montr\'{e}al, Canada, H3A 2T8}

\date{\today}

\begin{abstract}
Using diagrammatic methods, we show how the Ward identity can be used to constrain the ladder kernel in transport coefficient calculations.  More specifically, we use the Ward identity to determine the necessary diagrams that must be resummed using an integral equation.  One of our main results is an equation relating the kernel of the integral equation with functional derivatives of the full self-energy; it is similar to what is obtained with 2PI effective action methods.  However, since we use the Ward identity as our starting point, gauge invariance is preserved.  Using power counting arguments, we also show which self-energies must be included in the resummation at leading order, including 2 to 2 scatterings and 1 to 2 collinear scatterings with the LPM effect.  We show that our quantum field theory result is equivalent to the one of Arnold, Moore and Yaffe obtained using effective kinetic theory.  In this paper we restrict our discussion to electrical conductivity in hot QED, but our method can in principle be generalized to other transport coefficients and other theories.
\end{abstract}

\pacs{Valid PACS appear here}
\maketitle

\section{Introduction}
\label{sec:Intro}

Transport coefficients are measures of the efficiency at which a conserved quantity is transported on ``long'' distances (compared to microscopic relaxation scales) in a medium.  For example, electrical conductivity characterizes the diffusion of charge due to an external electric field and shear viscosity characterizes the diffusion of momentum transverse to the direction of propagation.  In real non-relativistic systems (e.g. interacting electron gas in a lattice), transport coefficients are almost impossible to calculate, because systems are strongly interacting and no simple closure exists; but in hot, weakly interacting theories, they can in principle be evaluated.  The computation of these quantities is important from a theoretical point of view: since they characterize linear deviations away from equilibrium but are computed from well-known equilibrium field theory tools, they could be used as a benchmark for testing non-equilibrium field theories, which are less well developed than equilibrium ones (but see the recent developments in Refs. \cite{Berges_Borsanyi_2005,Aarts_etal_2002,Berges_2002}).  They could also have an influence on the physics of the early universe, such as electroweak baryogenesis (see for example \cite{Rubakov_1996,Cohen_etal_1993}) and the formation and decay of primordial magnetic fields (see for example \cite{Turner_Widrow_1988,Giovannini_Shaposhnikov_2000,Boyanovsky_etal_2003a}).  Shear viscosity also attracted a lot of attention lately in the heavy ion community, partly due to the results on elliptic flow (seemingly implying a low viscosity, see for example \cite{Teaney_2003,PHENIX_2005} and the references therein) and its exact computation in a strongly coupled Super Yang-Mills theory \cite{Policastro_etal_2001}.  The above examples show the importance of having a good theoretical handle on transport coefficients.
  
The first calculation of transport coefficients in relativistic scalar theories can be found in \cite{Hosoya_etal_1984}.  Their calculation is based on Kubo relations, {\it i.e.} relations expressing transport coefficients in terms of retarded correlation functions between various conserved currents in the low momentum, low frequency limit.  The correlation functions are directly evaluated using finite temperature quantum field theory and thus provide a microscopic calculation of transport coefficients; but as is explained in \cite{Jeon_1995,Jeon_Yaffe_1996}, their calculation is incomplete.  Due to the use of resummed propagators (to regularize so-called pinch singularities), an infinite number of ladder diagrams must be resummed to get the leading order result.  This program has been carried out explicitly in \cite{Jeon_1995,Jeon_Yaffe_1996} for shear and bulk viscosities in scalar theories.  The shear viscosity result has since been reproduced using the real-time formalism \cite{Wang_Heinz_1999,Wang_Heinz_2003,Carrington_etal_2000}, direct ladder summation in Euclidean space \cite{Basagoiti_2002} and 2PI effective action methods \cite{Aarts_Martinez_2003}.  The results of both shear and bulk viscosities have been reproduced using quantum kinetic field theory derived from the closed-time-path 2PI effective action \cite{Calzetta_etal_2000}.

Order of magnitude estimates of transport coefficients in gauge theories based on phenomenology appeared more than 20 years ago in \cite{Danielewicz_1985}, but a complete leading order calculation in hot gauge theories just came out recently \cite{AMY_2000,AMY_2003a,AMY_2003b,Arnold_etal_2006}.  One of the main reason for this is the subtlety of the power counting involved, {\it i.e.} which scattering processes should be included at leading order.  In gauge theories, in addition to $2 \rightarrow 2$ scatterings with a soft momentum exchange, the presence of collinear singularities makes the $1 \rightarrow 2$ scatterings as important as the $2 \rightarrow 2$ ones \cite{Aurenche_etal_1996,Aurenche_etal_1997,Aurenche_etal_1998,AMY_2001}.  Moreover, interference effects between the various collinear emissions must also be taken into account at leading order; this is called the Landau-Pomeranchuk-Migdal (LPM) effect \cite{Landau_Pomeranchuk_1953,Migdal_1955,Aurenche_etal_2000,AMY_2001,Baym_etal_2006}.  The calculations in Refs. \cite{AMY_2000,AMY_2003a,AMY_2003b,Arnold_etal_2006} are based on kinetic theory and consistently include the physics of pinch singularities, collinear singularities and the LPM effect.  Let us mention that these calculations are also very involved technically, since the inclusion of pinch singularities implies the use of an integral equation and the inclusion of collinear singularities and the LPM effect implies the use of another integral equation embedded in the first one.

The equivalence between the quantum field theory approach and the kinetic theory approach in transport coefficients calculations was shown in \cite{Jeon_1995,Jeon_Yaffe_1996} in the case of scalar theories; but due to the complications mentioned previously, a similar equivalence has been lacking in gauge theories.  There exist some attempts at computing transport coefficients in hot gauge theories from quantum field theory using different approaches, such as direct ladder summation in Euclidean space \cite{Basagoiti_2002,Aarts_Martinez_2002,Defu_2005}, dynamical renormalization group methods \cite{Boyanovsky_etal_2003} and 2PI effective action methods \cite{Aarts_Martinez_2005}, but as far as we know, none of these approaches go beyond leading log order accuracy ({\it i.e.} with corrections suppressed by $O(g\ln(g^{-1}))$) or the large $N_{f}$ approximation.

The goal of this paper is to do a leading order calculation of transport coefficients in hot gauge theories using purely diagrammatic methods.  More precisely, we show the equivalence between quantum field theory and kinetic theory for transport coefficient calculations in hot gauge theories, thus justifying the effective kinetic theory results of Arnold, Moore and Yaffe \cite{AMY_2003b,AMY_2003a} from first principles.  This nontrivial check is reason enough for doing this calculation, but there are other reasons as well that go beyond kinetic theory.  For example, in cases where the configuration of the field itself is important (e.g. instanton) or unphysical particles with indefinite metric appear, then kinetic theory is inappropriate and one must rely on quantum field theory.  Also, quantum field theory might be the only way of computing transport coefficients beyond leading order; according to \cite{AMY_2003a}, doing the calculation using kinetic theory would require a whole new machinery.  It is not clear if calculations going beyond leading order can be converted to a linearized Boltzmann equation.  Finally, showing that the calculations are grounded and feasible in quantum field theory is important in itself.  It could give the necessary impetus for other quantum field theory methods (e.g. closed-time-path 2PI effective action \cite{Calzetta_etal_2000} or the dynamical renormalization group \cite{Boyanovsky_etal_2003}) to complete a leading order calculation and provide other insights into the problem.  In view of the applications mentioned previously, we think it is interesting and important to pursue this work.

In this paper, we restrict ourselves to Quantum Electrodynamics (QED) and electrical conductivity; we address the case of shear viscosity in a future paper \cite{Gagnon_Jeon_2006}.  The rest of the paper is organized as follows.  Section~\ref{sec:Background} presents our notation and some background material on transport coefficients, both in scalar and gauge theories.  Following the work of Ref.  \cite{Aarts_Martinez_2002}, Sect.~\ref{sec:Ward_identity} presents the derivation of the Ward identity in the limit appropriate for transport coefficient calculations.  It also presents the constraint on ladder kernels that can be obtained from the Ward identity.  Power counting arguments are shown in Sect.~\ref{sec:Power_counting}, in order to determine which rungs should be kept in the resummation.  The final expression for electrical conductivity, including collinear physics and the LPM effect, is presented in Sect.~\ref{sec:Integral_equations}.  We finally conclude in Sect.~\ref{sec:Conclusion}.  Technical details of some aspects of the calculations are relegated to the appendices.

\section{Background Material}
\label{sec:Background}

\subsection{Notation and Useful Formulas}
\label{sec:Notation}

We present here our notations and some useful formulas that are routinely used throughout the analysis.  Latin indices run from 1 to 3 and represent space components while Greek indices run from 0 to 3 and represent spacetime components.  Boldface, normal and capital letters denote 3-momenta, 4-momenta and Euclidean 4-momenta, respectively.  We use the metric convention $\eta_{\mu\nu} = (1,-1,-1,-1)$.  Sums over Matsubara frequencies are written as $\int\frac{d^{4}P}{(2\pi)^{4}} \equiv T\sum_{i\nu_{p}}\int\frac{d^{3}p}{(2\pi)^{3}}$, where $P = (i\nu_{p},\mathbf{p})$ and $\nu_{p} = 2n\pi T$ (bosons) or $\nu_{p} = (2n+1)\pi T$ (fermions) with $n$ an integer.  The subscripts $B$, $F$ attached to a quantity refer to its bosonic or fermionic nature (except for self-energies and widths, where we use a special notation).  The subscripts $R$, $I$ means real or imaginary part and the superscripts $\rm ret$, $\rm adv$, $\rm cor$ means retarded, advanced or autocorrelation (i.e. average value of the anti-commutator).  A bar over a quantity means that the gamma matrix structure is explicitly taken out (e.g. $G^{\mu}(k) \equiv \gamma^{\mu}\bar{G}(k)$).

We give a list of various finite temperature field theory quantities that are used in the rest of the paper.  We give their explicit expressions in momentum space and not their basic definitions in terms of fields (see for example \cite{Chou_etal_1985,Fetter_Walecka_2003}) because the latter are not useful for our purposes.  Free spectral densities are given by \cite{Jeon_Ellis_1998}:
\begin{eqnarray}
\label{eq:free_spectral_density_boson}
\rho_{B}(k) & = & \mbox{sgn}(k^{0})2\pi\delta((k^{0})^{2}-E_{k}^{2}) \\
\label{eq:free_spectral_density_fermion}
\rho_{F}(k) & = & \left[2\pi\delta(k^{0} - E_{k})h_{+}(\hat{k}) + 2\pi\delta(k^{0} + E_{k})h_{-}(\hat{k})\right]
\end{eqnarray}
where $E_{k} \equiv |\mathbf{k}|$, $h_{\pm}(\hat{k}) \equiv (\gamma^{0} \mp \mathbf{\gamma}\cdot\hat{k})/2$ and $\hat{k} \equiv \mathbf{k}/|\mathbf{k}|$.  Note that since we consider systems where the temperature is much larger than any other scale, we put $m = 0$ in the above and all subsequent expressions when the momentum of the excitation is hard.  From CPT, it can be shown that the spectral densities satisfy $\rho_{B}(-k^{0}) = -\rho_{B}(k^{0})$ and $\rho_{F}(-k) = \rho_{F}(k)$ (in the massless limit).  At finite temperature, any excitation propagating in a medium has a finite lifetime due to numerous collisions with on-shell thermal excitations.  The effect of this finite lifetime is to turn the delta functions in Eqs.~(\ref{eq:free_spectral_density_boson})-(\ref{eq:free_spectral_density_fermion}) into Lorentzians, giving \cite{Jeon_1995,Basagoiti_2002}:
\begin{eqnarray}
\label{eq:resumed_spectral_density_boson}
\rho_{B}(k) & = & \frac{1}{2E_{k}} \left[\frac{\gamma_{k}}{(k^{0}-E_{k})^{2}+(\gamma_{k}/2)^{2}}-\frac{\gamma_{k}}{(k^{0}+E_{k})^{2}+(\gamma_{k}/2)^{2}}\right] \\
\label{eq:resumed_spectral_density_fermion}
\rho_{F}(k) & = & \left[\frac{\Gamma_{k}}{(k^{0} - E_{k})^{2} + (\Gamma_{k}/2)^{2}}h_{+}(\hat{k}) + \frac{\Gamma_{k}}{(k^{0} + E_{k})^{2} + (\Gamma_{k}/2)^{2}}h_{-}(\hat{k})\right]
\end{eqnarray}
The widths are given by $\gamma_{k} \equiv \Pi_{I}^{\rm ret}(k^{0} = E_{k})/E_{k}$ and $\Gamma_{k} \equiv \mbox{tr}\left[k\!\!\!/ \Sigma_{I}^{\rm ret}(k^{0} = E_{k}) \right]/2E_{k}$, where $\Pi(k)$ and $\Sigma(k)$ are the bosonic and fermionic self-energies, respectively.  Note that when the momentum $k$ is soft, perturbation theory must be re-organized and partial resummation of spectral densities is necessary (also called Hard Thermal Loop (HTLs) resummations \cite{Braaten_Pisarski_1990,Taylor_Wong_1990,Braaten_Pisarski_1992}).  These resummations give rise to screening thermal masses and may also produce Landau damping.  In gauge theories, HTLs are also essential to obtain gauge invariant results (see Sect.~\ref{sec:Complications}).

The time-ordered (or ``uncut'') propagators can be expressed in terms of the spectral densities \cite{Jeon_Ellis_1998}:
\begin{eqnarray}
\label{eq:11_propagators}
G_{B/F}(k) & = & i\int\frac{d\omega}{(2\pi)}\; \rho_{B/F}(\omega)\left(\frac{1\pm n_{B/F}(\omega)}{k^{0}-\omega+i\epsilon} \pm \frac{n_{B/F}(\omega)}{k^{0}-\omega-i\epsilon} \right)
\end{eqnarray}
where $n_{B/F}(k^{0})$ are the usual Bose-Einstein or Fermi-Dirac distribution functions.  The anti time-ordered propagators are just the complex conjugate of the time-ordered ones.  Wightman (or ``cut'') propagators are given by \cite{Jeon_Ellis_1998}:
\begin{eqnarray}
\label{eq:12_propagators}
\Delta_{B/F}^{+}(k) & = & (1 \pm n_{B/F}(k^{0}))\rho_{B/F}(k) \\
\label{eq:21_propagators}
\Delta_{B/F}^{-}(k) & = & \pm n_{B/F}(k^{0})\rho_{B/F}(k)
\end{eqnarray}
The propagators~(\ref{eq:11_propagators})-(\ref{eq:21_propagators}) are the four propagators of the closed-time-path or ``1-2'' formalism \cite{Schwinger_1961,Keldysh_1965}, with the correspondence $G = G^{11}$, $G^{*} = G^{22}$, $\Delta^{+} = G^{12}$ and $\Delta^{-}  = G^{21}$.  Switching to the Keldysh (or $r$,$a$) basis, we can also write down the physical functions (see for example \cite{Chou_etal_1985}):
\begin{eqnarray}
\label{eq:physical_functions}
iG_{B/F}^{ra} \;\equiv\; iG_{B/F}^{\rm ret}(k) & = & G_{B/F}(k) - \Delta_{B/F}^{-}(k) \\
iG_{B/F}^{ar} \;\equiv\; iG_{B/F}^{\rm adv}(k) & = & G_{B/F}(k) - \Delta_{B/F}^{+}(k) \\
iG_{B/F}^{rr} \;\equiv\; iG_{B/F}^{\rm cor}(k) & = & \Delta_{B/F}^{+}(k) + \Delta_{B/F}^{-}(k)
\end{eqnarray}
The $G_{B/F}^{aa}$ is identically zero in the Keldysh basis.  Note also that any vertex in this basis must involve an odd number of $a$'s (see for example \cite{Gelis_1997}).  One can see that from the expression of the generating functional in the closed-time-path formalism, $Z = \int {\cal D}[\phi] \exp(i\int_{c}dt\int d^{3}x\;({\cal L}+J_{c}\phi))$.  Due to the integration over the closed-time-path, the Lagrangian is effectively ${\cal L} = {\cal L}(\phi_{1}) - {\cal L}(\phi_{2})$, where $\phi_{1}$ and $\phi_{2}$ are fields living on the time-ordered and anti time-ordered contours respectively.  From this we conclude that any interaction term is odd under the interchange of $\phi_{1}$ and $\phi_{2}$.  Switching to the Keldysh basis using $\phi_{r} = (\phi_{1}+\phi_{2})/2$  and $\phi_{a} = \phi_{1}-\phi_{2}$ , we see that any interaction must have an odd number of $a$'s, since only $\phi_{a}$ is odd under $\phi_{1} \leftrightarrow \phi_{2}$.  As a final remark, to get the explicit forms for free or resummed propagators, substitute in the appropriate spectral density (c.f. Eqs.~(\ref{eq:free_spectral_density_boson})-(\ref{eq:resumed_spectral_density_fermion})).

The cutting rules we use are the ones that can be found in \cite{Kobes_Semenoff_1985,Kobes_Semenoff_1986,Jeon_Ellis_1998} and are analogous to the zero-temperature ones.  The rules are:
\begin{enumerate}
	\item Draw all the cut diagrams relevant to the problem considered, where cuts separate the unshaded ({\it i.e.} ``1'') and the shaded ({\it i.e.} ``2'') regions.
	\item Use the usual Feynman rules for the unshaded region assigning $G_{B/F}(k)$ to the uncut lines.  For the shaded region, use the conjugate Feynman rules assigning $G_{B/F}^{*}(k)$ to the uncut lines.
	\item If the momentum of a cut line crosses from the unshaded to the shaded region, assign $\Delta_{B/F}^{+}(k)$.  If the momentum of a cut line crosses from the shaded to the unshaded region, assign $\Delta_{B/F}^{-}(k)$.
	\item Divide by the appropriate symmetry factor and multiply by an overall factor of $-i$.
\end{enumerate}
The various shadings given by the cutting rules are not all independent.  First, the ``vanishing of all circlings'' relation \cite{Kobes_Semenoff_1985,Kobes_Semenoff_1986,Jeon_Ellis_1998} says that the sum of all possible cuts of a given diagram is zero:
\begin{eqnarray}
\label{eq:Unitarity}
\sum_{a_{i} = 1,2} G_{B/F}^{a_{1}a_{2}...a_{n}} = 0
\end{eqnarray}
Equation~(\ref{eq:Unitarity}) can be rearranged so as to have the same form as a (generalized) optical theorem and is thus related to unitarity.  The vanishing of all circlings relation is only based on the observation that all propagators can be decomposed into a positive frequency part and a negative frequency part.  Second, the Kubo-Martin-Schwinger (KMS) relations express the proportionality between various pairs of shadings.  For example, for 2-point functions we have \cite{LeBellac_2000}:
\begin{eqnarray}
\label{eq:KMS_2pt_function}
G_{B/F}^{12}(k) & = & \pm e^{\beta k^{0}}G_{B/F}^{21}(k)
\end{eqnarray}
For 4-point functions, the relations are \cite{Carrington_etal_2000b,Wang_Heinz_2002}:
\begin{eqnarray}
\label{eq:KMS_4pt_function}
G_{B/F}^{*\; 2111}(k_{1},k_{2},k_{3},k_{4}) & = & \pm e^{-\beta k_{1}^{0}} G_{B/F}^{1222}(k_{1},k_{2},k_{3},k_{4}) \nonumber \\
G_{B/F}^{*\; 1211}(k_{1},k_{2},k_{3},k_{4}) & = & \pm e^{-\beta k_{2}^{0}} G_{B/F}^{2122}(k_{1},k_{2},k_{3},k_{4}) \nonumber \\
G_{B/F}^{*\; 1121}(k_{1},k_{2},k_{3},k_{4}) & = & \pm e^{-\beta k_{3}^{0}} G_{B/F}^{2212}(k_{1},k_{2},k_{3},k_{4}) \nonumber \\
G_{B/F}^{*\; 1112}(k_{1},k_{2},k_{3},k_{4}) & = & \pm e^{-\beta k_{4}^{0}} G_{B/F}^{2221}(k_{1},k_{2},k_{3},k_{4}) \nonumber \\
G_{B/F}^{*\; 2211}(k_{1},k_{2},k_{3},k_{4}) & = & e^{-\beta (k_{1}^{0}+k_{2}^{0})} G_{B/F}^{1122}(k_{1},k_{2},k_{3},k_{4}) \nonumber \\
G_{B/F}^{*\; 2121}(k_{1},k_{2},k_{3},k_{4}) & = & e^{-\beta (k_{1}^{0}+k_{3}^{0})} G_{B/F}^{1212}(k_{1},k_{2},k_{3},k_{4}) \nonumber \\
G_{B/F}^{*\; 2112}(k_{1},k_{2},k_{3},k_{4}) & = & e^{-\beta (k_{1}^{0}+k_{4}^{0})} G_{B/F}^{1221}(k_{1},k_{2},k_{3},k_{4})
\end{eqnarray}
where $\beta$ is the inverse temperature and energy-momentum requires $k_{1}+k_{2}+k_{3}+k_{4} = 0$.  The relations~(\ref{eq:KMS_2pt_function})-(\ref{eq:KMS_4pt_function}) are only based on the (anti) periodicity in imaginary times of the Green's functions and are thus only valid in equilibrium.

\subsection{Transport Coefficients in Relativistic Scalar Field Theory}
\label{sec:Transport_scalar}

A natural starting point for evaluating transport coefficients using quantum field theory is a Kubo type relation.  These relations express transport coefficients in terms of long distance correlations between conserved currents.  Restricting ourselves to electrical conductivity ($\sigma$), we have \cite{Hosoya_etal_1984,Jeon_1995,Jeon_Yaffe_1996,Basagoiti_2002,AMY_2000,Kapusta_Gale_2006}:
\begin{eqnarray}
\label{eq:Kubo_relations}
\sigma & = & \frac{\beta}{6} \lim_{k^{0} \rightarrow 0,\;{\bf k}=0} \int d^{4}x\; e^{i k\cdot x}\; \langle j_{i}(t,{\bf x})j^{i}(0) \rangle_{\rm eq}
\end{eqnarray}
where $j^{\mu}(x)$ is the electric current density.  Other Kubo relations exist for shear and bulk viscosity \cite{Hosoya_etal_1984,Jeon_1995,Jeon_Yaffe_1996,Basagoiti_2002,AMY_2000}.  Equivalently, one can express these Kubo relations in terms of derivatives of the spectral density, itself equal to twice the imaginary part of the corresponding retarded correlator.  Note that the averages are done with respect to an equilibrium density matrix, even though transport coefficients are non-equilibrium quantities.  

To compute transport coefficients using these relations, one needs to expand the retarded 2-point function perturbatively.  As explained in details in Refs. \cite{Jeon_1995,Jeon_Yaffe_1996}, an infinite number of ladder diagrams contribute to the transport coefficient, even at lowest order.  The reason for this is the low frequency limit in the Kubo relation.  This limit gives rise to products of propagators $G(p)$ with the same momentum (see Fig.~\ref{fig:ladder_diagrams}).  Since finite temperature propagators possess four poles (one in each quadrant), one faces situations when the integration contour is ``pinched'' between two poles on opposite sides of the real axis in the complex $p^{0}$ plane .  In equations, we have ($E_{p}$ is the on-shell quasi-particle energy):
\begin{figure}
\resizebox{5in}{!}{\includegraphics{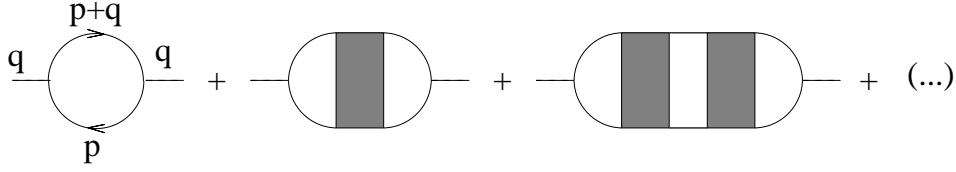}}
\caption{\label{fig:ladder_diagrams} Examples of ladder diagrams, where the grey squares represent 4-point functions called ``rungs''.  When the external momentum $q$ goes to zero, the two ``side rail'' propagators have the same momentum and produce a ``pinch'' singularity.}
\end{figure}
\begin{eqnarray}
\label{eq:Pinch_singularities}
\int\frac{d^{0}p}{(2\pi)}\; G_{B}(p)G_{B}(p) & \sim & \int\frac{d^{0}p}{(2\pi)}\; G_{B}^{\rm ret}(p)G_{B}^{\rm adv}(p) \nonumber \\
                                             & \sim & \int\frac{d^{0}p}{(2\pi)}\; \left(\frac{1}{(p^{0}+i\epsilon)-E_{p}}\right) \left(\frac{1}{(p^{0}-i\epsilon)-E_{p}}\right) \;\; \sim \;\; \frac{1}{\epsilon}
\end{eqnarray}
which diverges when $\epsilon$ goes to zero.  This divergence is symptomatic of the infinite lifetime of the excitation.  In a medium, excitations constantly suffer collisions from on-shell excitations coming from the thermal bath, resulting in a finite lifetime.  Thus, the use of resummed propagators regularizes these ``pinch'' singularities, effectively replacing $1/\epsilon$ with $1/\Pi_{I}(p)$.  The explicit coupling constants that now appear in the denominator change the power counting dramatically and make the resummation of an infinite number of ladder diagrams necessary \cite{Jeon_1995,Jeon_Yaffe_1996}.

This resummation is done by re-writing the infinite sum of ladder diagrams in terms of an effective vertex, itself solution to an integral equation.  Schematically, we have for electrical conductivity:
\begin{eqnarray}
\label{eq:Kubo_relation_schematic}
\sigma & = & \frac{\beta}{6} \int\frac{d^{4}k}{(2\pi)^{4}}\; {\cal I}_{F}^{*}(k){\cal F}(k){\cal D}_{F}(k)
\end{eqnarray}
%
\begin{figure}
\resizebox{5in}{!}{\includegraphics{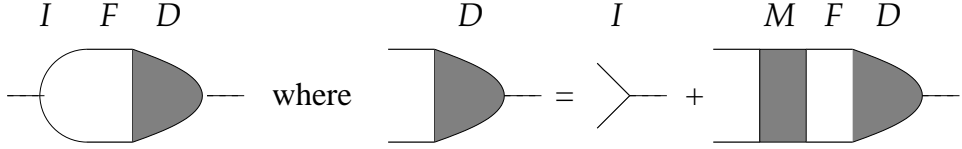}}
\caption{\label{fig:integral_equation} Schematic representation of Eqs.~(\ref{eq:Kubo_relation_schematic})-(\ref{eq:integral_equation_schematic}).  The symbols refer directly to the equations: ${\cal I}$ is an external current insertion, ${\cal M}$ is a rung, ${\cal F}$ represents a pair of side rail propagators and ${\cal D}$ is an effective vertex.}
\end{figure}
\begin{eqnarray}
\label{eq:integral_equation_schematic}
{\cal D}_{F}(k) & = & {\cal I}_{F}(k) + \int\frac{d^{4}p}{(2\pi)^{4}}\; {\cal K}(k,p){\cal D}_{F}(p)
\end{eqnarray}
where ${\cal K} \equiv {\cal M}{\cal F}$.  See Fig.~\ref{fig:integral_equation} for a graphical representation of Eqs.~(\ref{eq:Kubo_relation_schematic})-(\ref{eq:integral_equation_schematic}).  The symbol ${\cal F}$ represents a pair of ``side rail'' propagators (note that the ``ladder'' diagram in Fig.~\ref{fig:integral_equation} is on its side, meaning that the ``side rails'' are on the top and bottom of the diagram).  In the limit $q\rightarrow 0$, the two side rail propagators that hook up ${\cal M}$ to ${\cal D}_{F}$ have the same momentum and ``pinch'', producing a $1/\Sigma_{I}(p)$ factor.  The ``rungs'' ${\cal M}$ are 4-point functions that must be of the same order as ${\cal F}^{-1}$ but otherwise arbitrary.  The effective vertex ${\cal D}$ encodes the information about the infinite resummation of ladder diagrams.  If we close it with an external current insertion ${\cal I}$ (with an additional factor of ${\cal F}$ to connect them), we get the Kubo relation~(\ref{eq:Kubo_relation_schematic}).  Note that because of the ${\cal F}$ factor, the transport coefficient gets a non-analytic behavior in the coupling constant.  This is expected, since transport coefficients are roughly proportional to the mean free path and thus inversely proportional to the scattering cross section of the processes responsible for transport.

To write down the appropriate integral equation~(\ref{eq:integral_equation_schematic}) and solve it are the main tasks of transport coefficient calculations.  Altough it can become very involved numerically (especially for gauge theories), there is no conceptual problem with solving the integral equation.  On the other hand, to know which rungs contribute at leading order is difficult and requires detailed power counting arguments.  This is the approach adopted in \cite{Jeon_1995}, where they find that a certain (finite) set of rungs are necessary and sufficient to compute the shear and bulk viscosities at leading order in a $g\phi^{3}+\lambda\phi^{4}$ theory.

\subsection{Additional Complications in Gauge Theories}
\label{sec:Complications}

Gauge theories are considered to be the most successful theories on the market to describe the fundamental interactions of Nature.  So for ``real'' applications (e.g. viscous hydrodynamic evolution of the QGP), the need to extend transport coefficient calculations to gauge theories is obvious.  From experience, we know that gauge theories are more complicated than scalar theories, and the present calculations are no exceptions.  We make a list of the main complications that arise when computing transport coefficients in hot gauge theories.

Preserving gauge invariance is an important problem when dealing with gauge theories.  In practice, Ward identities tell us how non-gauge invariant quantities (such as propagators and vertices) must be related to each other so as to preserve gauge invariance.  We show in Sect.~\ref{sec:Ward_identity} how Ward identities are used to this effect.

Another complication, not specific to gauge theories, is related to the presence of different species of particles (electrons, photons, etc) and particles with different statistics.  The fact that there are fermions in the theory means that the tools developed in \cite{Jeon_1995} must be slightly modified to take into account the fermionic nature of the particles.  These complications are not major and rather technical in nature.  We show in Sect.~\ref{sec:Integral_equations} how the tools in \cite{Jeon_1995} must be modified in the presence of these complications.

The distinction between hard ($O(T)$) and soft ($O(eT)$) momenta is also  particularly important in hot gauge theories ($e$ is the electromagnetic coupling constant).  In particular, it is shown in \cite{Braaten_Pisarski_1990,Taylor_Wong_1990,Braaten_Pisarski_1992} that the theory must be partially resummed when soft momenta are present to get gauge invariant results.  These Hard Thermal Loop (HTL) resummations give rise to screening thermal masses and modify the form of the propagators (vertices are also modified but this not necessary for our purposes).  For numerical purposes (as in \cite{AMY_2003a,AMY_2003b}), the expressions of HTL-resummed propagators are required.  In the present paper, our goal is to reproduce the Boltzmann equation of Arnold, Moore and Yaffe \cite{AMY_2003a,AMY_2003b} and we never use the explicit form of the propagators; the calculations of Sect.~\ref{sec:Integral_equations} are based on very general properties that do not depend on HTL resummation, such as even/odd properties of spectral densities, KMS conditions and unitarity.  In this sense, this technical complication does not concern us.

But surely the most important complication comes from the fact that, in gauge theories, transport coefficients are sensitive to soft and collinear physics even at leading order \cite{AMY_2003a,AMY_2003b}.  In scalar theories, explicit power counting shows that soft momenta and collinear singularities do not have any effect \cite{Jeon_1995}.  As with pinch singularities, collinear singularities come from the multiplication of two propagators with momenta $p$ and $p+q$ that are nearly collinear ({\it i.e.} $\mathbf{p}\cdot\mathbf{q} \sim O(e^{2}T^{2})$):
\begin{eqnarray}
\int\frac{d^{0}p}{(2\pi)}\; G_{B}(p)G_{B}(p+q) & \sim & \int\frac{d^{0}p}{(2\pi)}\; G_{B}^{\rm ret}(p)G_{B}^{\rm adv}(p+q) \nonumber \\
                                             & \sim & \int\frac{d^{0}p}{(2\pi)}\; \left(\frac{1}{(p^{0}+i\epsilon)-E_{p}}\right) \left(\frac{1}{(p^{0}+q^{0}-i\epsilon)-E_{p+q}}\right) \nonumber \\
					     & \sim & \frac{1}{q^{0}+(E_{p+q}-E_{p}) - 2i\epsilon}
\end{eqnarray}
In the limit $q \rightarrow 0$ (pinch singularities), the expression diverges as $1/\epsilon$ and we get back Eq.~(\ref{eq:Pinch_singularities}).  In the case where $q$ is nearly on-shell but nonzero and the angle between the quasi-particles is small ({\it i.e.} $\theta_{pq} \sim e$), we have $E_{p+q} \approx E_{p} \pm |\mathbf{q}|$ and the integral diverges as $1/\theta_{pq}^{2}$ (or $1/\epsilon$ in the perfectly collinear and massless case).  These collinear singularities must be regulated by including a finite width in the propagators.  As with the pinch singularity case, the introduction of coupling constants in the denominator changes the power counting dramatically.  This collinear enhancement affects an infinite class of diagrams: this is the well-known LPM effect (see for example \cite{Aurenche_etal_2000,AMY_2001,Baym_etal_2006} for a discussion of this effect in the context of photon production).  Because there are now two types of singularities, there are two resummations to do.  While in scalar theories there is an infinite number of ladder diagrams and a finite number of different rung types, in gauge theories there are an infinite number of ladder diagrams (due to pinch singularities) and an infinite number of different rung types (due to collinear singularities).  We come back to these issues and the subtleties of power counting in gauge theories in Sect.~\ref{sec:Power_counting}.

\section{Ward Identity Constraints on the Ladder Kernels}
\label{sec:Ward_identity}

\subsection{Ward Identities for Charge}
\label{sec:Derivation_Ward}

As mentioned in Sect.~\ref{sec:Complications}, the task of finding which rungs contribute at leading order is not easy.  The purpose of the present section and the next one is to show that there exists a guiding principle that can help us find the structure of the rungs.  

A necessary piece of information in our demonstration is the Ward identity for the effective vertex, as derived for the first time for the case of an electric current insertion in Ref. \cite{Aarts_Martinez_2002}.  We show here the essential steps.  The starting point is the usual Ward identity for charge conservation in Euclidean space
\begin{figure}
\resizebox{2in}{!}{\includegraphics{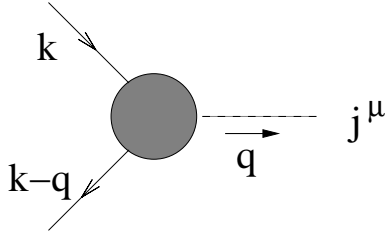}}
\caption{\label{fig:Ward_identity} Momentum convention for the Ward identity (Eq.~(\ref{eq:Ward_charge_Euclidean})).  The dotted line stands for an insertion of an electric current and the blob represents the amputated fermionic effective vertex ${\cal D}_{F}$.}
\end{figure}
\begin{eqnarray}
\label{eq:Ward_charge_Euclidean}
Q_{\mu}{\cal D}_{F}^{\mu}(K,K-Q) & = & G_{F}^{-1}(K) - G_{F}^{-1}(K-Q) 
\end{eqnarray}
where ${\cal D}_{F}$ is a fermionic amputated effective vertex.  The momentum convention for Eq.~(\ref{eq:Ward_charge_Euclidean}) is shown in Fig.~\ref{fig:Ward_identity}.  To go from Euclidean space to Minkowski space, we need to analytically continue $K$ and $Q$ towards real energies.  The proper choice here is dictated by the physics of pinch singularities and the Kubo relation.  To have pinch singularities, one needs the multiplication of two propagators with different boundary conditions, namely $G_{F}^{\rm ret}(k)G_{F}^{\rm adv}(k-q)$ or $G_{F}^{\rm adv}(k)G_{F}^{\rm ret}(k-q)$; the Kubo relation requires the evaluation of a retarded current-current 2-point function.  These two requirements uniquely fix the analytic continuation to $K \rightarrow k^{0} + i\epsilon$, $K-Q \rightarrow k^{0} + q^{0} - i\epsilon$ and $Q \rightarrow q^{0} + 2i\epsilon$.  With this analytic continuation, Eq.~(\ref{eq:Ward_charge_Euclidean}) becomes
\begin{eqnarray}
\label{eq:Ward_charge_Minkowski}
q_{\mu}{\cal D}_{F}^{\mu}(k + i\epsilon,k-q - i\epsilon) & = & G_{F\;\rm ret}^{-1}(k) - G_{F\;\rm adv}^{-1}(k-q)
\end{eqnarray}
Taking the necessary limits $q^{0} \rightarrow 0$ and ${\bf q} \rightarrow 0$ (c.f. Eq.~(\ref{eq:Kubo_relations})) and using $G_{F\;\rm ret/adv}^{-1}(p) = [\gamma^{0}(p^{0}\pm i\Gamma_{p}/2) - {\bf\gamma} \cdot {\bf p}]$ (valid when $p$ is nearly on-shell), we get
\begin{eqnarray}
\label{eq:Ward_charge}
\lim_{q \rightarrow 0}\; q_{\mu}{\cal D}_{F}^{\mu}(k + i\epsilon,k-q - i\epsilon) & = & i\gamma^{0}\Gamma_{k} \;=\; 2i\Sigma_{I}^{\rm ret}({\bf k})
\end{eqnarray}
The last equality is valid near $k^{0}\approx \mathbf{k}$.  This last equation relates the effective vertex of the integral equation to the imaginary part of the on-shell retarded self-energy in the limit relevant to transport in the case of an electric current insertion.  Note that we used resummed retarted/advanced propagators because of the need to regularize pinch singularities.  As a final remark, note that it is possible to repeat the above derivation for a $T^{\mu\nu}$ insertion \cite{Gagnon_Jeon_2006}.  This is particularly important when calculating shear viscosity from Kubo relations.

\subsection{Derivation of the Constraint}
\label{sec:Derivation_constraint}

In this section, we show that there exists a relation between the kernel of the integral equation~(\ref{eq:integral_equation_schematic}) and the imaginary part of the on-shell self-energy.  The ladder kernel being a 4-point function and the self-energy being a 2-point function, the relation between the two must necessarily involve functional derivatives with respect to propagators ({\it i.e.} ``opening'' of lines in a Feynman diagram).  Such relations have been obtained using 2PI effective action methods \cite{Aarts_Martinez_2003,Aarts_Martinez_2005}.  Here we derive similar constraints but starting from a more physical point of view, namely the Ward identities of the previous section.

In the following, we restrict ourselves to electrical conductivity ({\it i.e.} only one integral equation is involved), but the discussion can be generalized to viscosity with some effort (although the result is not as clean cut as for the electrical conductivity) \cite{Gagnon_Jeon_2006}.  Starting from the integral equation~(\ref{eq:integral_equation_schematic}) in Euclidean space,
\begin{eqnarray}
{\cal D}_{F}^{\mu}(K,K-Q)  & = & {\cal I}_{F}^{\mu}(K,K-Q) + \int\frac{d^{4}P}{(2\pi)^{4}}{\cal K}(K,P,Q){\cal D}_{F}^{\mu}(P,P-Q)
\end{eqnarray}
our goal is to isolate the kernel and express it in terms of known quantities.  Multiplying both sides by $Q_{\mu}$ and using Eq.~(\ref{eq:Ward_charge_Euclidean}), we get
\begin{eqnarray}
\label{eq:Integral_equation_Euclidean_modified}
Q_{\mu}{\cal D}_{F}^{\mu}(K,K-Q)  & = & Q_{\mu}{\cal I}_{F}^{\mu}(K,K-Q) + \int\frac{d^{4}P}{(2\pi)^{4}}{\cal M}(K,P,Q)[G_{F}(P-Q)-G_{F}(P)]
\end{eqnarray}
where we have separated the kernel into a ``rung'' part and a ``side rail'' part, ${\cal K}(K,P,Q) = {\cal M}(K,P,Q)G_{F}(K)G_{F}(K-P)$.  Note that this is a schematic notation and it does not reflect correctly the Dirac matrix structure of the integral equation; one must be aware of this fact when doing explicit calculations.  To make further progress, it is necessary to analytically continue towards real energies.  This step is delicate, since the sum over Matsubara frequencies must be done first.  As explained in more detail in Appendix~\ref{app:Analytic_continuation}, the Matsubara sum can be done using the summation formula found in \cite{Basagoiti_2002}, with ${\cal M}$ replaced by the spectral representation of a general 4-point function.  The physics of the Kubo relation and pinch singularities can then be implemented by doing the same analytic continuation as for the Ward identities (see Sect.~\ref{sec:Derivation_Ward}).  The result is that the integral equation keeps its form in Minkowski space and is similar to the results of Jeon \cite{Jeon_1995} (suitably generalized to fermions).  Taking the $q\rightarrow 0$ limit and using the Ward identity~(\ref{eq:Ward_charge}), we get
\begin{eqnarray}
\label{eq:Integral_equation_Minkowski}
2i\Sigma_{I}^{\rm ret}(k)  & = & \int\frac{d^{4}p}{(2\pi)^{4}}{\cal M}(k,p)[G_{F}^{\rm adv}(p)-G_{F}^{\rm ret}(p)]
\end{eqnarray}
with the understanding that $k$ and $p$ are on-shell in the pinching pole limit.  Using the definition of the spectral density $\rho_{F}^{+}(p) \equiv i(G_{F}^{\rm ret}(p)-G_{F}^{\rm adv}(p))$, we can rewrite the integral equation as
\begin{eqnarray}
\label{eq:constraint_1}
2\Sigma_{I}^{\rm ret}(k)  & = & \int\frac{d^{4}p}{(2\pi)^{4}}{\cal M}(k,p) \rho_{F}^{+}(p)
\end{eqnarray}
Here $\Sigma_{I}^{\rm ret}(k)$ is the full (imaginary) self-energy and ${\cal M}$ is the coherent addition of all possible rungs.  Since $\Sigma_{I}^{\rm ret}$ and $\rho_{F}^{+}$ are known quantities, Eq.~(\ref{eq:constraint_1}) expresses a constraint on the rung part of the ladder kernel ${\cal M}$.  To get a more useful expression, we need to invert Eq.~(\ref{eq:constraint_1}).  Let us functionally differentiate both sides with respect to $\rho_{F}^{+}(q)$
\begin{eqnarray}
\label{eq:Integral_equation_first_derivative}
\frac{\delta \left(2\Sigma_{I}^{\rm ret}(k)\right)}{\delta \rho_{F}^{+}(q)} & = & \int\frac{d^{4}p}{(2\pi)^{4}} \left(\frac{\delta {\cal M}(k,p)}{\delta \rho_{F}^{+}(q)}\right) \rho_{F}^{+}(p) + {\cal M}(k,q)
\end{eqnarray}
All cut/uncut propagators can be expressed in terms of the spectral density (c.f. Eqs.~(\ref{eq:11_propagators})-(\ref{eq:21_propagators})), thus differentiating with respect to $\rho_{F}^{+}(q)$ can be interpreted as the opening of a fermion line in a Feynman diagram.  Such a diagrammatic interpretation is of course not perfect, since $\rho_{F}^{+}$ is only proportional to fermionic propagators.  Some distribution functions are left behind when interpreting Eq.~(\ref{eq:Integral_equation_first_derivative}) diagrammatically, but since we are only interested in the diagrammatic structure of ${\cal M}$, a proportionality relation is sufficient for our purposes.

Equation~(\ref{eq:Integral_equation_first_derivative}) can be further reduced.  The key observation is that ${\cal M}$ is an addition of Feynman diagrams made of an arbitrary number of fermion propagators.  The following relation
\begin{eqnarray}
\label{eq:Inverse_operation}
\int\frac{d^{4}p}{(2\pi)^{4}}\left(\frac{\delta {\cal A}(k,q)}{\delta \rho_{F}^{+}(p)}\right)\rho_{F}^{+}(p) & = & \alpha{\cal A}(k,q)
\end{eqnarray}
is valid for any Feynman diagram ${\cal A}$ (in particular for ${\cal M}$ and $\Sigma_{I}^{\rm ret}$), where $\alpha$ is the number of fermion propagators in ${\cal A}$.  In words, it means that the ``inverse operator'' of a functional derivative with respect to a fermion propagator is an integration and a multiplication by a fermion propagator (with the same momentum present in the function we differentiate with).  It is easy to see that Eq.~(\ref{eq:Inverse_operation}) is true in simple examples (both for bosons and fermions).

Equation~(\ref{eq:Inverse_operation}) cannot be applied directly to Eq.~(\ref{eq:Integral_equation_first_derivative}), because the momenta in the functional derivative and in the spectral density are not the same.  In order to apply the inverse operator~(\ref{eq:Inverse_operation}) and thus use the special condition that ${\cal M}$ is a Feynman diagram, we need to manipulate Eq.~(\ref{eq:Integral_equation_first_derivative}).  Let us functionally differentiate Eq.~(\ref{eq:Integral_equation_first_derivative}) a second time, giving
\begin{eqnarray}
\label{eq:Integral_equation_second_derivative}
\frac{\delta {\cal M}(k,q)}{\delta \rho_{F}^{+}(r)} + \frac{\delta {\cal M}(k,r)}{\delta \rho_{F}^{+}(q)} & = & -\int\frac{d^{4}p}{(2\pi)^{4}} \left(\frac{\delta^{2} {\cal M}(k,p)}{\delta \rho_{F}^{+}(r) \delta \rho_{F}^{+}(q)}\right) \rho_{F}^{+}(p) + \frac{\delta^{2} \left(2\Sigma_{I}^{\rm ret}(k)\right)}{\delta \rho_{F}^{+}(r) \delta \rho_{F}^{+}(q)}
\end{eqnarray}
Note that this last equation expresses the ``symmetrized'' first order functional derivative of the rung kernel in terms of second functional derivatives of the rung kernel and self-energy.  Plugging Eq.~(\ref{eq:Integral_equation_second_derivative}) back into Eq.~(\ref{eq:Integral_equation_first_derivative}), we get
\begin{eqnarray}
\label{eq:Integral_equation_first_derivative_modified}
\frac{\delta \left(2\Sigma_{I}^{\rm ret}(k)\right)}{\delta \rho_{F}^{+}(q)} & = & - \int\frac{d^{4}p}{(2\pi)^{4}} \left(\frac{\delta {\cal M}(k,q)}{\delta \rho_{F}^{+}(p)}\right)\rho_{F}^{+}(p) + \int\frac{d^{4}p}{(2\pi)^{4}} \left(\frac{\delta^{2} \left(2\Sigma_{I}^{\rm ret}(k)\right)}{\delta \rho_{F}^{+}(p) \delta \rho_{F}^{+}(q)}\right)\rho_{F}^{+}(p)  \nonumber \\
	   &   & -\int\frac{d^{4}p}{(2\pi)^{4}}\int\frac{d^{4}l}{(2\pi)^{4}} \left(\frac{\delta^{2} {\cal M}(k,l)}{\delta \rho_{F}^{+}(p) \delta \rho_{F}^{+}(q)}\right) \rho_{F}^{+}(l)\rho_{F}^{+}(p)  + {\cal M}(k,q)
\end{eqnarray}
Written in this form, Eq.~(\ref{eq:Integral_equation_first_derivative_modified}) can be simplified using the inverse operator.  For convenience, we introduce the following notation
\begin{eqnarray}
\label{eq:Decomposition_cut_propagators}
{\cal M} & = & \sum_{\alpha = 1}^{\infty} {\cal M}_{\alpha} \nonumber \\
\Sigma_{I}^{\rm ret} & = & \sum_{\beta = 1}^{\infty} \Sigma_{I\;\beta}^{\rm ret}
\end{eqnarray}
where the sums over $\alpha$ and $\beta$ are over the number of fermion propagators (note that the ${\cal M}_{\alpha}$'s or $\Sigma_{I\;\beta}^{R}$ can be sums of diagrams in themselves).  These definitions mean that, in general, ${\cal M}$ and $\Sigma_{I}^{\rm ret}$ are ``blobs'' containing diagrams of all orders; these diagrams can be re-organized in terms of their number of fermion propagators.  Applying Eqs.~(\ref{eq:Inverse_operation}) and (\ref{eq:Decomposition_cut_propagators}) to Eq.~(\ref{eq:Integral_equation_first_derivative_modified}) and doing some algebra, we obtain
\begin{equation}
\label{eq:Constraint_derivative_M}
\sum_{\alpha = 1}^{\infty} \alpha \left[ \int\frac{d^{4}l}{(2\pi)^{4}} \left( \frac{\delta {\cal M}_{\alpha}(k,l)}{\delta \rho_{F}^{+}(q)} \right)\rho_{F}^{+}(l) + \frac{(1-\alpha)}{\alpha}\left(\frac{\delta (2\Sigma_{I\;\alpha}^{\rm ret}(k))}{\delta \rho_{F}^{+}(q)}\right) - \frac{(1-\alpha)}{\alpha} {\cal M}_{\alpha}(k,q) \right]  =  0
\end{equation}
This last equation is valid component by component, {\it i.e.} for each value of the number of fermion propagators $\alpha$.  This is a constraint equation, implementing the condition that ${\cal M}$ is a Feynman diagram and the use of the inverse operator.  Plugging Eq.~(\ref{eq:Constraint_derivative_M}) back into the first integral equation~(\ref{eq:Integral_equation_first_derivative}) and using Eq.~(\ref{eq:Decomposition_cut_propagators}), we get
\begin{eqnarray}
\label{eq:Constraint_M_charge}
{\cal M}_{\alpha}(k,q) = \left(\frac{\delta (2\Sigma_{I\;\alpha}^{\rm ret}(k))}{\delta \rho_{F}^{+}(q)}\right) & \rightarrow & {\cal M}(k,q) = \left(\frac{\delta (2\Sigma_{I}^{\rm ret}(k))}{\delta \rho_{F}^{+}(q)}\right)
\end{eqnarray}
valid for all values of $\alpha$.  The second equality is obtained by summing over $\alpha$ on both sides and using Eqs.~\ref{eq:Decomposition_cut_propagators}.  This is the desired constraint on the ladder kernel.  Note that it is possible to obtain relations similar to Eq.~(\ref{eq:Constraint_M_charge}) for shear viscosity \cite{Gagnon_Jeon_2006}.  So even if we restrict ourselves to the case of electrical conductivity in QED in this paper, our method is general and can in principle be applied to other transport coefficients and other theories.

\subsection{Interpretation of the Constraint}
\label{sec:Interpretation_constraint}

As pointed out in Sect.~\ref{sec:Transport_scalar}-\ref{sec:Complications}, one of the hardest tasks in computing transport coefficients by diagrammatic methods is to find the right rung kernel in Eq.~(\ref{eq:integral_equation_schematic}).  In other words, the ultimate goal is to find all the appropriate rungs that must be included in ${\cal M}$ so as to obtain the transport coefficient at the desired level of accuracy.  The usual procedure is to consider all possible rungs cut in all possible ways and use power counting arguments to show which ones contribute at, say, leading order.  This way to proceed is already involved in scalar theories; in gauge theories, it would quickly become hard to manage. 

\begin{figure}
\resizebox{\textwidth}{!}{\includegraphics{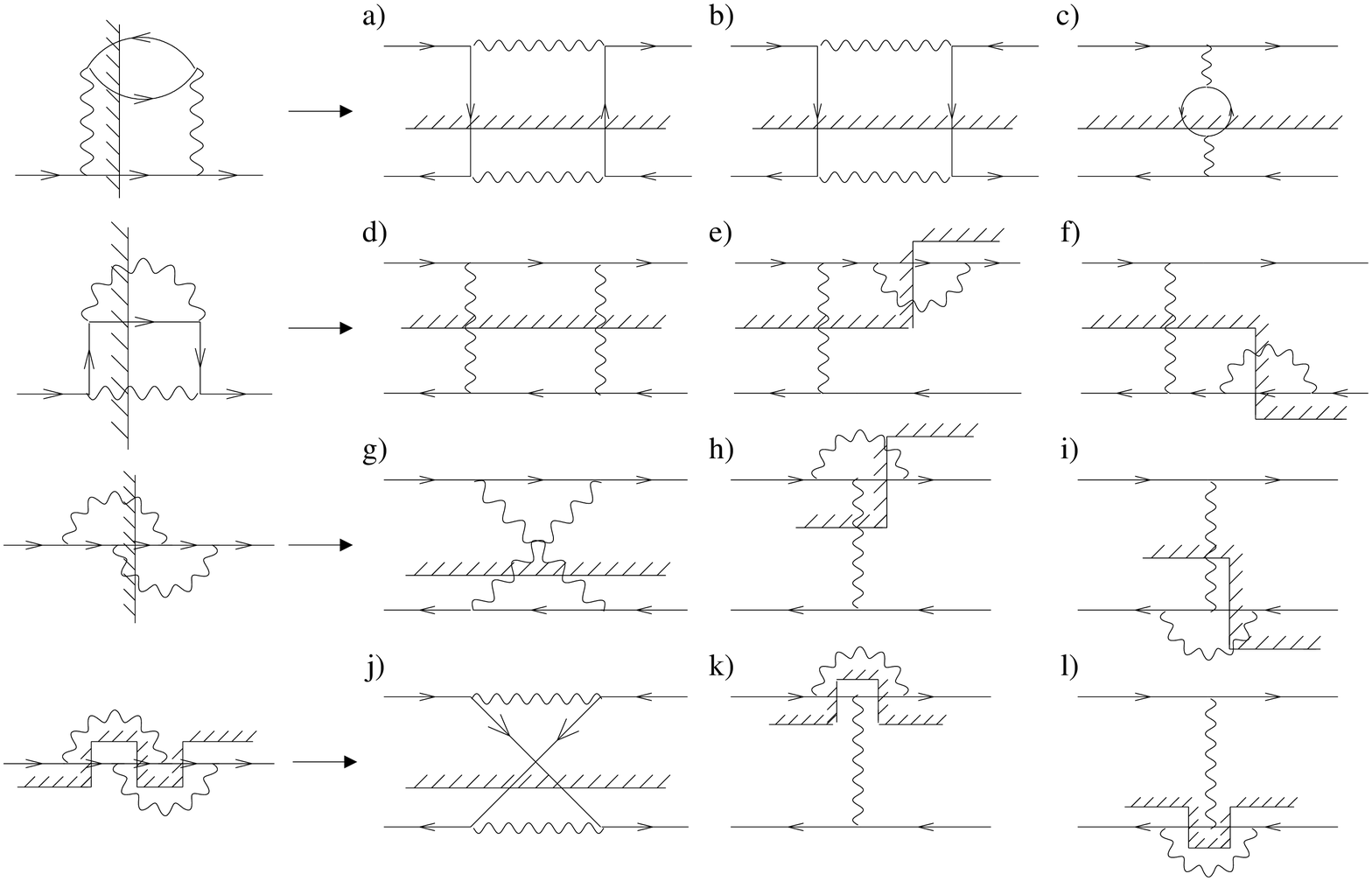}}
\caption{\label{fig:2to2_rungs} Rungs obtained by functionnally differentiating with respect to a fermionic spectral density the self-energies in the left column (c.f. Eq.~\ref{eq:Constraint_M_charge}).  Power counting arguments show that all these rungs contribute at leading order except diagrams (e) and (f), because massless three-body on-shell decays are suppressed.  All diagrams can be converted into $2\rightarrow 2$ scatterings with a soft exchange except (h) and (i), which are part of the $1\rightarrow 2$ collinear scatterings discussed in \cite{AMY_2003a,AMY_2003b,AMY_2001}.}
\end{figure}

Equation~(\ref{eq:Constraint_M_charge}) provides a way to restrict the number of possible cut rungs that must be included in ${\cal M}$.  It says that the full rung kernel is proportional to the functional derivative of the imaginary part of the full retarded on-shell self-energy.  This result means that any rung kernel (a 4-point function) is obtained by opening a fermionic line in an imaginary self-energy (a 2-point function).  The recipe to find ${\cal M}$ is thus to select a certain set of imaginary self-energies (using for example an expansion in coupling constants) and open the fermion lines in these self-energies to get the rungs.  An example of such a procedure applied to one-loop self-energies in QED is shown in Fig.~\ref{fig:2to2_rungs}.  Since we are dealing with a quantum mechanical system, all such openings must be summed coherently.  This is a strong constraint on the rung kernel.

A few remarks are in order here.  The above constraint implements the physics of transport coefficients ({\it i.e.} pinch singularities due to the low frequency, low momentum limit) and is useful for restricting the number of possible cut rungs: it is a necessary condition to get leading order rungs.  In this case, power counting is used to verify if the rungs obtained from the constraint contribute at leading order.  For electrical conductivity, this verification is done in Sect.~\ref{sec:Power_counting_without_collinear}.  If other singularities are present, then the constraint is not sufficient and must be supplemented with additional power counting arguments.  This is the case in gauge theories, where collinear singularities make rungs with number changing processes as important as those without these processes.  These additional power counting arguments are presented in Sect.~\ref{sec:Power_counting_with_collinear}.

The constraint equation~(\ref{eq:Constraint_M_charge}) expresses rungs in terms of functional derivatives of self-energies.  Since a few self-energies can give rise to many rungs, it could be more convenient to work directly on the self-energies in some cases.  This is particularly relevant when, because of the presence of collinear singularities, an infinite number of rungs are needed at leading order.  Such an analysis could rapidly become unmanageable; as shown in Sect.~\ref{sec:Power_counting_with_collinear}, using power counting to carefully select the relevant self-energies before opening them makes the analysis easier. 


The constraint~(\ref{eq:Constraint_M_charge}) is also essential to preserve gauge invariance in our formalism.  To see that more clearly, let us compare our formalism to the 2PI formalism.  Expressions similar to Eq.~(\ref{eq:Constraint_M_charge}) were obtained using 2PI effective actions methods \cite{Aarts_Martinez_2003,Aarts_Martinez_2005}, with notable differences.  In 2PI methods, the ``constraint'' is in coordinate space and comes naturally from standard functional relations.  The important point is that the kernel ${\cal M}_{2PI}$ of the integral equation is given by the functional derivative of the self-energy with respect to resummed propagators, where the self-energy is itself given by the functional derivative of all amputated 2PI diagrams ($\Gamma_{2}$) with respect to resummed propagators:
\begin{eqnarray}
\label{eq:Constraint_2PI}
{\cal M}_{2PI}(x,y;a,b)  & = & 2\frac{\delta\Pi(x,y)}{\delta G_{B}(a,b)} \;=\; 2\frac{\delta}{\delta G(a,b)}\left(2i\frac{\delta \Gamma_{2}}{\delta G_{B}(x,y)}\right) 
\end{eqnarray}
To obtain Eq.~(\ref{eq:Constraint_2PI}), no reference is made to the low frequency, low momentum limit or pinch singularities.  In that sense, the ``constraint'' in \cite{Aarts_Martinez_2003,Aarts_Martinez_2005} is too general for the problem at hand.  We also mention that there seem to be gauge invariance issues with 2PI methods \cite{Arrizabalaga_Smit_2002,Mottola_2003,Carrington_etal_2005}.  Briefly, this is because the 2PI effective action methods involve the use of resummed propagators, {\it i.e.} only a certain class of topologies are resummed.  On the other hand, gauge theories require Ward identities to be satisfied.  Since Ward identities require cancellations between different classes of topologies (resummed propagators and vertices), they are necessarily violated in the 2PI formalism.  Note that it is shown in \cite{Berges_2004} that a complete, self-consistent system of equations for the dynamics of QED or QCD can be obained using higher order effective actions, but it has never been implemented in practice for transport coefficients.

In contrast, our method does not make any reference to 2PI effective actions and starts directly from symmetry principles, the Ward identity being the expression of these symmetries for quantum mechanical amplitudes.  The physical limit that is used to obtain transport coefficients ($q \rightarrow 0$) is explicitly implemented in our constraint (Eq.~(\ref{eq:Constraint_M_charge})).  In addition, the constraint contains, by construction, exactly those diagrams required to produce a gauge invariant result for the electrical conductivity in QED; in other words, it tells us which self-energy must be resummed in the side rail propagators for each rung present in the kernel so as to keep everything transverse.  This constraint is not as general as the constraint obtained from 2PI methods (compare Eqs.~(\ref{eq:Constraint_M_charge}) and (\ref{eq:Constraint_2PI})) or from general considerations of gauge invariance, but it is more powerful in the sense that it contains the relevant physics to calculate the electrical conductivity in QED.

\section{Power Counting}
\label{sec:Power_counting}

Equation~(\ref{eq:Constraint_M_charge}) gives us an infinite number of self-energies from which rungs can be obtained by ``opening'' lines.  To make progress, we need a selection criterion based on power counting to isolate the rungs that contribute at leading order.  Naively, one could do an expansion in coupling constants and keep only the lowest order diagrams, but this procedure turns out not to be sufficient.  As pointed out in Sect.~\ref{sec:Complications}, due to the presence of collinear singularities, there is a restricted but infinite class of diagrams that must be resummed to get leading order results.  The goal of this section is to present power counting arguments to obtain the leading order electric conductivity.  For the purpose of this section, we divide the rungs into two categories: the ones containing collinear singularities and the ones that do not.  In kinetic theory language, these correspond to $1 + N \rightarrow 2 + N$ and $2\rightarrow 2$ scatterings, respectively (we come back to this last point in Sect.~\ref{sec:Integral_equations}).

\subsection{Power Counting Without Collinear Singularity}
\label{sec:Power_counting_without_collinear}

We first consider the case of no collinear singularity.  According to Eq.~(\ref{eq:Constraint_M_charge}), the rungs are obtained by opening fermion lines in the imaginary part of the retared self-energy of the electron.  At one loop, the imaginary part is zero since an on-shell massless excitation cannot decay into two on-shell massless excitations.  It is thus necessary to go to two loops for a leading order result.  The two-loop imaginary retarded self-energies are shown in Fig.~\ref{fig:2to2_rungs}, with their corresponding rungs.  Note that there are many more cuts that correspond to imaginary two-loop self-energies.  We consider these other cuts when writing down the $4\times 4$ matrix integral equation (c.f.~(\ref{eq:4x4_integral_equation})); for the moment, we are only interested in the rung topology.  Let us check the power counting size of each rung.

\begin{figure}
\resizebox{4in}{!}{\includegraphics{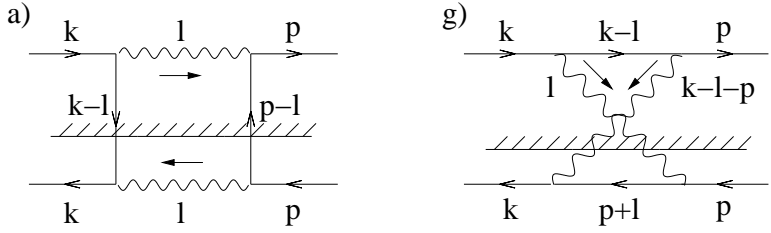}}
\caption{\label{fig:diagram_a_g} Momentum labels used to do the power counting of rungs (a) and (g) (c.f. Fig.~\ref{fig:2to2_rungs}).}
\end{figure}

Consider rung (a) in Fig.~\ref{fig:2to2_rungs}, reproduced in Fig.~\ref{fig:diagram_a_g} with momentum labels.  The expression for the rung is:
\begin{eqnarray}
\label{eq:Rung_a}
{\cal M}_{(a)} & = & -ie^{4}\int\frac{d^{4}l}{(2\pi)^{4}} \Delta_{F}^{-}(k-l)\Delta_{F}^{+}(p-l) G_{B}^{*}(l)G_{B}(l)
\end{eqnarray}
where for clarity we omitted Dirac matrices and Lorentz indices (irrelevant for non-collinear power counting).  If all momenta are hard, then the rung is $O(e^{4})$ (this is true for all rungs considered here).  In the case of a soft bosonic exchange $l \sim O(eT)$, the story is different.  Each bosonic propagator is now $O(e^{-2}T^{-2})$ (note that there is no contribution from the Bose-Einstein distribution since $l$ is off-shell) and there is a phase space suppression of $e^{2}$ ({\it i.e.} two integrals are killed by the delta functions inside the cut propagators, leaving $d^{2}l \sim O(e^{2}T^{2})$).  Collecting all powers, we get that rung (a) is $O(e^{2})$ for a soft exchange.  This is the usual Coulomb divergence in scattering theory and is the dominant part of the integral.  Rungs (b) and (c) also contain a Coulomb divergence; the power counting is done in the same way and gives $O(e^{2})$.

The structure of rung (d) is very similar to rung (a), with the role of bosons and fermions interchanged.  The only difference in the power counting comes from the fact that the exchanged soft excitation is now a fermion.  Since a (soft) fermion propagator is $O(e^{-1}T^{-1})$ (instead of $O(e^{-2}T^{-2})$ for a soft boson propagator), rung (d) is suppressed by two additional powers of $e$ compared to rung (a), giving $O(e^{4})$.  Rungs (e) and (f) are necessarily higher order than the others and should not be considered.  This is due to the presence of three-point vertices with their three legs on-shell and massless, corresponding to kinematically forbidden or highly suppressed processes.

The power counting of rung (g) in Fig.~\ref{fig:2to2_rungs} is slightly more involved.  The expression for the rung is (the momentum convention is shown in Fig.~\ref{fig:diagram_a_g}):
\begin{eqnarray}
\label{eq:Rung_g}
\int\frac{d^{4}p}{(2\pi)^{4}}\;{\cal M}_{(g)} & = & -ie^{4}\int\frac{d^{4}p}{(2\pi)^{4}}\int\frac{d^{4}l}{(2\pi)^{4}} \Delta_{B}^{-}(l)\Delta_{B}^{-}(k-l-p) G_{F}^{*}(k-l)G_{F}(p+l)
\end{eqnarray}
where we explicitly write the integration over $p$ coming from the integral equation (c.f. Eq.~(\ref{eq:integral_equation_schematic})).  We consider the case where the side rail momenta $k$ and $p$ are both hard and nearly on-shell ({\it i.e.} $k^{2}\sim p^{2}\sim O(e^{2}T^{2})$) while the loop momentum $l$ is soft.  In such a kinematical regime, the dominant contribution comes from when the propagor momenta satisfy $(k-l)^{2} \sim (p+l)^{2} \sim O(e^{2}T^{2})$.  For these conditions to be satisfied, the angles $\theta_{kl}$ and $\theta_{pl}$ must be $O(e)$, implying an $e^{2}$ suppression in both phase space integrations (this is because $d^{4}p = dp^{0}\,|p|^{2}\sin\theta\, d|p|\,d\theta\,d\phi \sim O(e^{2})$ when $\theta \sim O(e)$).  Another way to see the phase space suppression in $d^{4}l$ is to notice that the two delta functions inside the cut propagators kill two integrals over $l$, leaving only $d^{2}l \sim O(e^{2})$ when $l$ is soft.  Accounting for all powers of $e$, we have an $e^{4}$ coming from the four explicit vertices, $e^{2}\times e^{2}$ from the phase space integrations and $e^{-2}\times e^{-2}$ coming from the two fermion propagators (note that the fermion propagators are not $O(e^{-1}T^{-1})$ because their upstair momenta are hard).  Rung (g) is thus $O(e^{4})$.  The power counting and size of rungs (j), (k) and (l) are similar to rung (g).  Finally, rungs (h) and (i) are suppressed except in the very special kinematical regime where the electron is collinear to the exchanged photon; they are part of the ``collinearly singular'' rungs   and will be considered shortly.

From the above power counting, we get that the rungs with a Coulomb divergence ({\it i.e.} (a), (b), (c)) are $O(e^{2})$ and the rungs without this divergence ({\it i.e.} (d), (g),  (j), (k), (l)) are $O(e^{4})$.  Naively, one would expect only rungs (a)-(c) to contribute at leading order, but it turns out that there exists another suppression mechanism that is only effective for these rungs.  To understand this suppression, let's look at the $2\rightarrow 2$ collision term of the linearized Boltzmann equation of Arnold, Moore and Yaffe \cite{AMY_2000} (we show in Sect.~\ref{sec:Integral_equations} how to obtain this equation, c.f. Eq.~(\ref{eq:Integral_equation_reduced_5})):
\begin{eqnarray}
C^{u}({\bf k}) & = & \frac{1}{2}\sum_{v,m,n}^{f,s,h}\int\frac{d^{3}p}{(2\pi)^{3}2E_{p}}\frac{d^{3}l}{(2\pi)^{3}2E_{l}}\frac{d^{3}l'}{(2\pi)^{3}2E_{l'}}\; |M_{uvmn}(k,p;l,l')|^{2} (2\pi)^{4}\delta^{4}(k+p-l-l') \nonumber \\
               &   & \times\; n^{u}(k)n^{v}(p)(1\pm n^{m}(l))(1\pm n^{n}(l')) \left[ \chi^{u}({\bf k}) + \chi^{v}({\bf p}) - \chi^{m}({\bf l}) - \chi^{n}({\bf l'}) \right]
\end{eqnarray}
where the sum is over flavors ($f$), species ($s$) and helicities ($h$), the $M_{uvmn}$'s are the $2\rightarrow 2$ scatterings relevant for the calculation and the $\chi$'s represent small deviations away from equilibrium distribution functions.  The key observation to understand the suppression is that rungs (a), (b) and (c) correspond to $2\rightarrow 2$ processes for which both incident excitations undergo a soft scattering (${\bf k}-{\bf p} = {\bf q}$, where ${\bf q}$ is $O(eT)$) without changing their species types ($u = m$ and $v = n$); see Fig.~\ref{fig:2to2_scatterings} for the correspondence between rungs and scattering processes.  In such a case, there is a partial cancellation between the $\chi$'s in the collision term \cite{AMY_2000}:
\begin{figure}
\resizebox{5in}{!}{\includegraphics{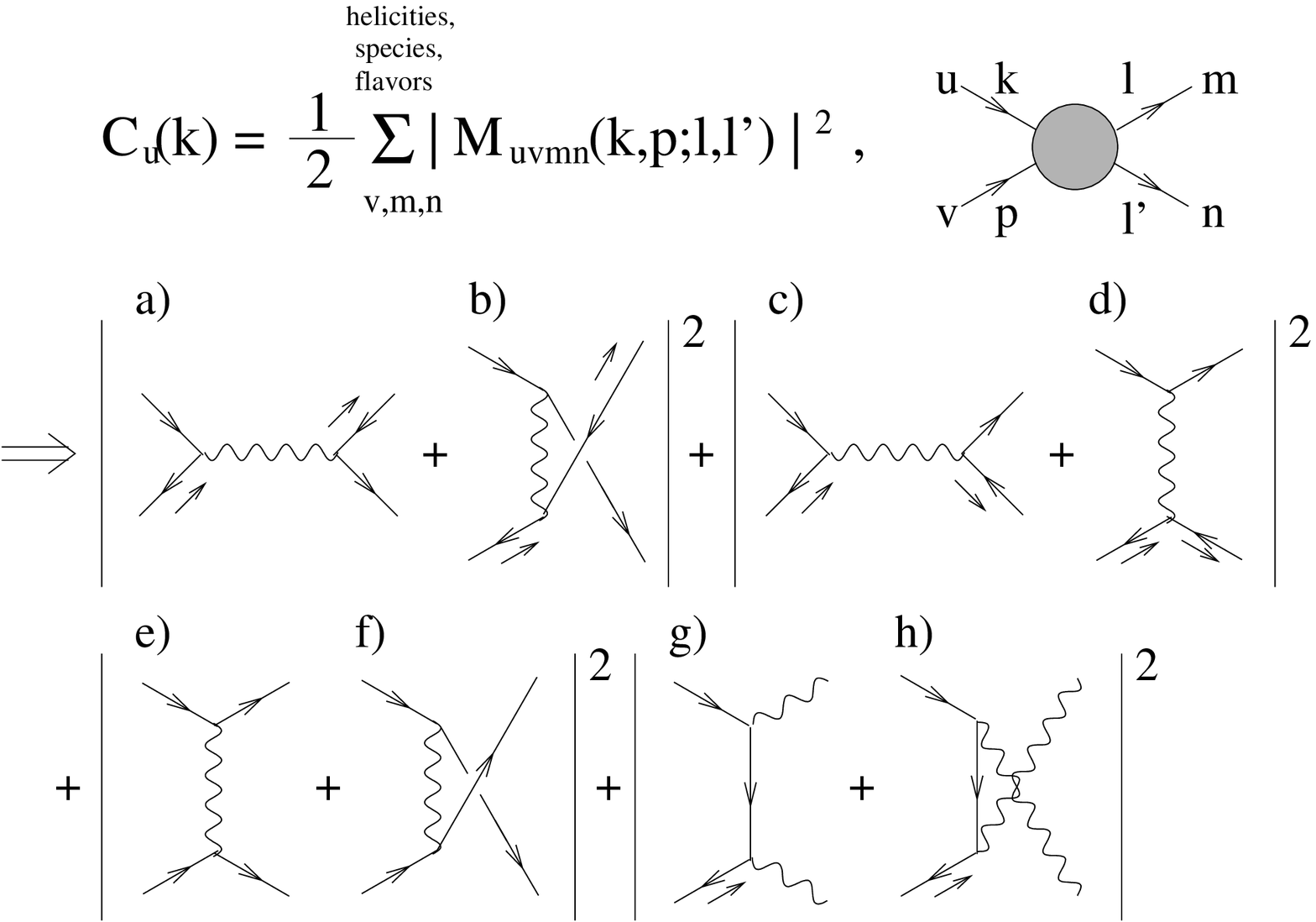}}
\caption{\label{fig:2to2_scatterings} Collision term of the linearized Boltzmann equation including only leading order $2\rightarrow 2$ scatterings \cite{AMY_2003a,AMY_2003b}.  The labels $u$-$n$ represent the species (fermion or photon), flavor and helicity of the excitations (here $u =$ electron).  A tedious calculation shows the equivalence between the scattering processes and the rungs of Fig.~\ref{fig:2to2_rungs}.  The correspondence goes as follows (the letters refer to diagrams in Figs.~\ref{fig:2to2_rungs} and \ref{fig:2to2_scatterings}, respectively): (a) $=$ (b)$^{2}$ and (d)$^{2}$, (b) $=$ (e)$^{2}$ and (f)$^{2}$, (c) $=$ (a)$^{2}$ and (c)$^{2}$, (d) $=$ (g)$^{2}$ and (h)$^{2}$,(g) $=$ (g)(h), (j) $=$ (e)(f), (i) $=$ (a)(b) or (c)(d), (j) $=$ (c)(d) or (a)(b).}
\end{figure}
\begin{eqnarray}
\chi^{u}({\bf k}) - \chi^{u}({\bf p}) & = & -{\bf q}\cdot \nabla\chi^{u}({\bf k}) + O(q^{2}) \;\; \sim \;\; O(eT) + O(q^{2})
\end{eqnarray}
Since in the Kubo relation~(\ref{eq:Kubo_relation_schematic}) the term $\left[ \chi^{u}({\bf k}) + \chi^{v}({\bf p}) - \chi^{m}({\bf l}) - \chi^{n}({\bf l'}) \right]$ is squared, the overall suppression is $e^{2}$.  Taking into account this additional suppression, rungs (a), (b) and (c) are $O(e^{4})$.

In summary, rungs (a)-(c), (d), (g) and (j)-(l) in Fig.~\ref{fig:2to2_rungs} all contribute at leading order for electric conductivity, although the contribution of rungs (a) and (b) are effectively zero due to Furry's theorem.  This completes the power counting for leading order rungs corresponding to $2\rightarrow 2$ scattering processes in QED.  Note that the Ward-like identity constraint is also applicable to scalar theories; it can easily be shown that taking all the two-loop imaginary self-energies of $g\phi^{3}+\lambda\phi^{4}$ gives the necessary leading order rungs \cite{Jeon_1995}.  We can conjecture that it is necessary and sufficient to use the Ward identity constraint and an expansion in coupling constants to get all rungs of a given accuracy that do not depend on another singularity (such as collinear singularities).  On the other hand, the constraint only implements the physics of pinch singularities and thus cannot give precise information about rungs containing collinear singularities.

\subsection{Power Counting With Collinear Singularities}
\label{sec:Power_counting_with_collinear}

Let us now come back to the problem of collinear singularities.  As explained in Sect.~\ref{sec:Complications}, collinear singularities appear when two propagators $G(k)$ and $G(k+p)$ nearly pinch, {\it i.e.} when the angle between $k$ and $p$ is $\theta_{kp}\sim O(e)$.  Thus we must look for cut rungs containing number changing processes and pairs of nearly pinching propagators.  By inspection, we find that the only rung topologies satisfying those criteria are the ones with vertex corrections, such as rungs (h) and (i) in Fig.~\ref{fig:2to2_rungs}.

Let us check the power counting size of rung (h), reproduced with momentum labels in Fig.~\ref{fig:diagram_h} (rung (i) is done in a similar way).  The expression corresponding to rung (h) is:
\begin{figure}
\resizebox{2.5in}{!}{\includegraphics{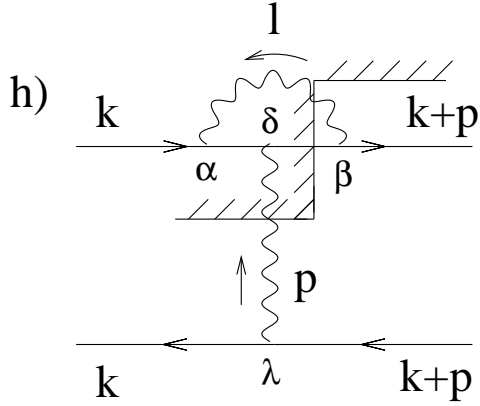}}
\caption{\label{fig:diagram_h} Momentum labels used to do the power counting of rung (h) (c.f. Fig.~\ref{fig:2to2_rungs}).}
\end{figure}
\begin{eqnarray}
\label{eq:power_counting_h_1}
\int\frac{d^{4}p}{(2\pi)^{4}}\;{\cal M}_{(h)}{\cal D}_{F}(p) & = & -ie^{4}\int\frac{d^{4}p}{(2\pi)^{4}}\int\frac{d^{4}l}{(2\pi)^{4}} \Delta_{B\;\alpha\beta}^{+}(l)\Delta_{B\;\delta\lambda}^{+}(p) \nonumber \\
                                                             &   & \times \gamma^{\alpha} G_{F}^{*}(k+l)\gamma^{\delta}\Delta_{F}^{-}(k+p+l) \gamma^{\beta}{\cal D}_{F}(p)\gamma^{\lambda}
\end{eqnarray}
where we explicitly write the integration over $p$ and the effective vertex coming from the integral equation (c.f. Eq.~(\ref{eq:integral_equation_schematic})).  The integral over $dl^{0}$ is dominated by the kinematical range where the two fermionic propagators $k+l$ and $k+p+l$ nearly pinch.  In such a regime, expression~(\ref{eq:power_counting_h_1}) becomes:
\begin{eqnarray}
\label{eq:power_counting_h}
\int\frac{d^{4}p}{(2\pi)^{4}}\;{\cal M}_{(h)}{\cal D}_{F}(p) & \sim & -ie^{4}\int\frac{d^{4}p}{(2\pi)^{4}}\int\frac{d^{3}l}{(2\pi)^{3}} \Delta_{B\;\alpha\beta}^{+}(l)\Delta_{B\;\delta\lambda}^{+}(p) \nonumber \\
               &      & \times \gamma^{\alpha} \left(\frac{(k\!\!\!/+l\!\!\!/)\gamma^{\delta}(k\!\!\!/+p\!\!\!/+l\!\!\!/)}{p^{0}+(E_{k+l}-E_{k+l+p})-\frac{i}{2}(\Gamma_{k+l}+\Gamma_{k+p+l})}\right) \gamma^{\beta}{\cal D}_{F}(p)\gamma^{\lambda}
\end{eqnarray}
The external legs are hard and nearly on-shell due to pinch singularities, {\it i.e.} $k\sim (k+p)\sim T$ and $k^{2}\sim (k+p)^{2}\sim O(e^{2}T^{2})$.  We also assume that the exchange photon $p$ is hard (otherwise the rung would be subleading due to too much phase space suppression); it is also on-shell since it is cut.  The nearly pinching conditions are given by $(k+l)^{2}\sim O(e^{2}T^{2})$ and $(k+p+l)^{2}\sim O(e^{2}T^{2})$.  This is equivalent to the statement that $l^{2}$, $(k\cdot l)$ and $(p\cdot l)$ must be $O(e^{2}T^{2})$ or that the electron on the upper rail must be collinear with the exchange photon.  In particular, it implies that $\theta_{pl}\sim O(e)$, thus restricting the phase space of $p$ to an $O(e^{2})$ region since $d^{4}p = dp^{0}\,|p|^{2}\sin\theta\, d|p|\,d\theta\,d\phi \sim O(e^{2})$ when $\theta \sim O(e)$.

Up to now, we have not specified the vertex correction momentum $l$: it can be either hard or soft.  Let us first consider the case when $l$ is hard.  In this momentum regime, the power counting size of rung (h) is $e^{4}\times e^{2}\times e^{2}\times [\Delta_{B\;\alpha\beta}^{+}(l)]\times [\gamma^{\alpha}(k\!\!\!/+l\!\!\!/)\gamma^{\delta}(k\!\!\!/+p\!\!\!/+l\!\!\!/)\gamma^{\beta}]\times e^{-2}$, where the $e^{4}$ comes from the four explicit vertices, $e^{2}\times e^{2}$ from the (small angle) restriction on the phase space of $p$ and $l$, and $e^{-2}$ from the nearly pinching propagators.  Since $l$ is hard, the bosonic cut propagator is $O(e^{0})$ and is proportional to $g_{\alpha\beta}$.  In this case, the Dirac structure reduces to a scalar product and generates an additional $e^{2}$ suppression.  Collecting all powers of $e$, we get that rung (h) is $O(e^{8})$ when $l$ is hard.

The situation is different when $l$ is soft.  In this case, the power counting size of rung (h) is $e^{4}\times e^{2}\times e^{3}\times [\Delta_{B\;\alpha\beta}^{+}(l)]\times [\gamma^{\alpha}(k\!\!\!/+l\!\!\!/)\gamma^{\delta}(k\!\!\!/+p\!\!\!/+l\!\!\!/)\gamma^{\beta}]\times e^{-2}$, where the $e^{4}$ comes from the four explicit vertices, $e^{2}$ from the restriction on the phase space of $p$, $e^{3}$ from the soft integration over $l$ and $e^{-2}$ from the nearly pinching propagators.  Since $l$ is soft, the cut propagator is HTL resummed and is no longer proportional to $g_{\alpha\beta}$.  There is thus no additional $e^{2}$ suppression coming from the Dirac structure when $l$ is soft.  The power counting size of $\Delta_{B\;\rm HTL}^{+}(l) \sim (1+n_{B}(l^{0}))\rho_{B\;\rm HTL}(l)$ is also variable and depends on the momentum flowing through it.  For soft spacelike momenta, Landau damping gives rise to an $O(e^{2})$ imaginary self-energy.  In this situation, $\rho_{B\;\rm HTL}(l)\sim \Sigma_{I}(l)/(l^{2}+\Sigma_{I}(l))^{2} \sim e^{-2}$ and $\Delta_{B\;\rm HTL}^{+}(l) \sim e^{-3}$.  Collecting all powers, we find that rung (h) is $O(e^{4})$ in the collinear regime.  On the other hand, for null or soft timelike momenta, $\Sigma_{I}(l)\sim O(e^{3})$ (no Landau damping) and $\rho_{B\;\rm HTL}(l)\sim e^{-1}$, making the rung subleading.

\begin{figure}
\resizebox{\textwidth}{!}{\includegraphics{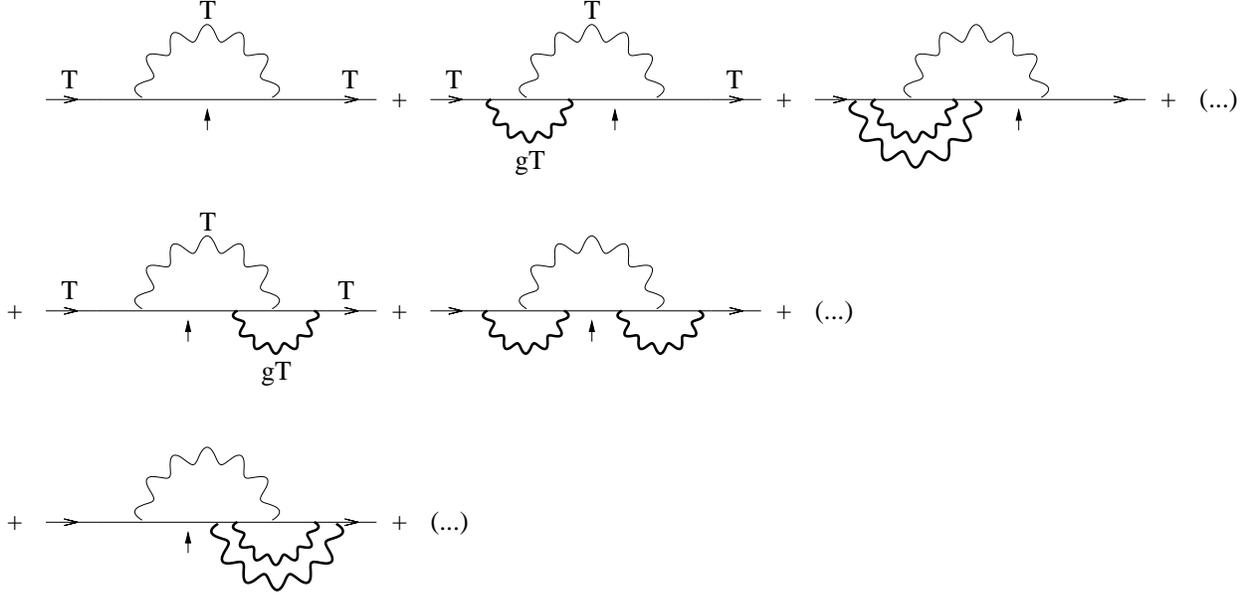}}
\caption{\label{fig:1to2_self_energies_conductivity} Self-energies considered in our leading order electrical conductivity calculation.  Soft photons are indicated by thick lines; all other excitations are hard.  In each diagram, the hard incoming electron is collinear to the hard photon.  Any number of soft vertex corrections must be included due to the presence of nearly pinching pairs of progagators; this is the diagrammatic depiction of the LPM effect.  The propagators indicated by vertical arrows are the only one that can be opened without breaking a pair of nearly pinching propagators.}
\end{figure}

\begin{figure}
\resizebox{\textwidth}{!}{\includegraphics{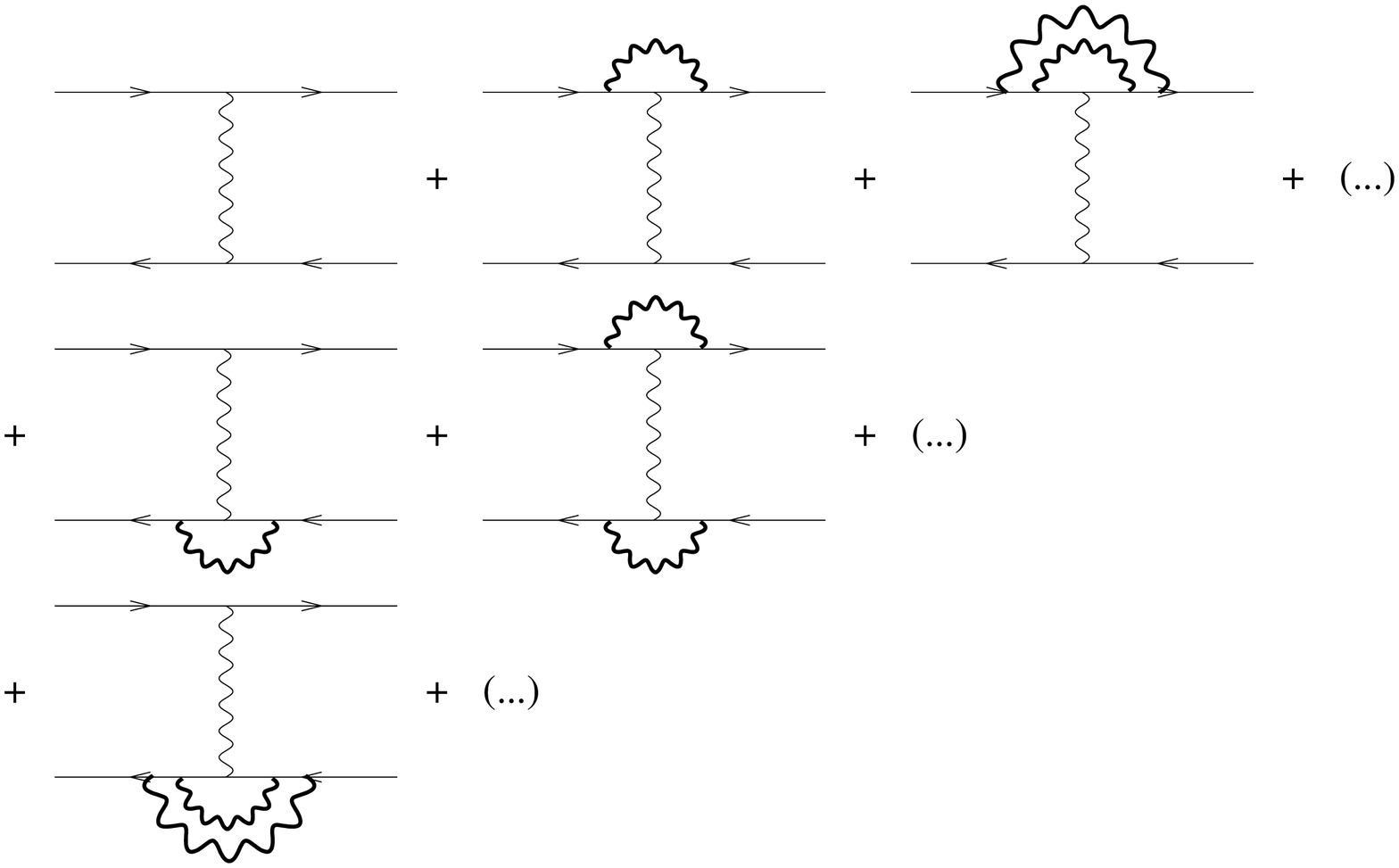}}
\caption{\label{fig:1to2_rungs_conductivity} Possible rung topologiess containing $1\rightarrow 2$ collinear processes obtained by opening (allowed) fermion lines in the self-energies of Fig.~\ref{fig:1to2_self_energies_conductivity}.  Soft photons are indicated by thick lines; all other excitations are hard.  The leading order rungs are in the first row and column; all rungs with at least one self-energy correction on both the upper and lower rails are subleading because of the impossibility of having upper and lower pairs of propagators pinch at the same time.}
\end{figure}

More generally, rungs obtained by opening lines on self-energies with any number of vertex corrections (as shown in Fig.~\ref{fig:1to2_self_energies_conductivity}) could be leading order (see below for an explicit example of power counting with two self-energy corrections).  The only restriction is that one should not open a pinching propagator; since pinch singularities occur for pairs of propagators, opening (removing) one of them just cancels the effect and would thus make the rung subleading.  With this restriction in mind, opening the self-energies in Fig.~\ref{fig:1to2_self_energies_conductivity} results in the rungs of Fig.~\ref{fig:1to2_rungs_conductivity}.  Before going further, let us verify the power counting size of the rung in Fig.~\ref{fig:diagram_h} with an additional vertex correction either on the upper or lower rail.  It modifies Eq.~(\ref{eq:power_counting_h_1}) in the following way: it adds two explicit vertices ($e^{2}$), one integral over soft 3-momenta ($e^{3}$), two pinching propagators ($e^{-2}$) and one cut propagator with soft spacelike momentum ($e^{-3}$).  Collecting together the newly introduced powers of $e$, we see that adding a vertex correction results in an $O(1)$ correction.  This is true for any number of vertex corrections.

\begin{figure}
\resizebox{5in}{!}{\includegraphics{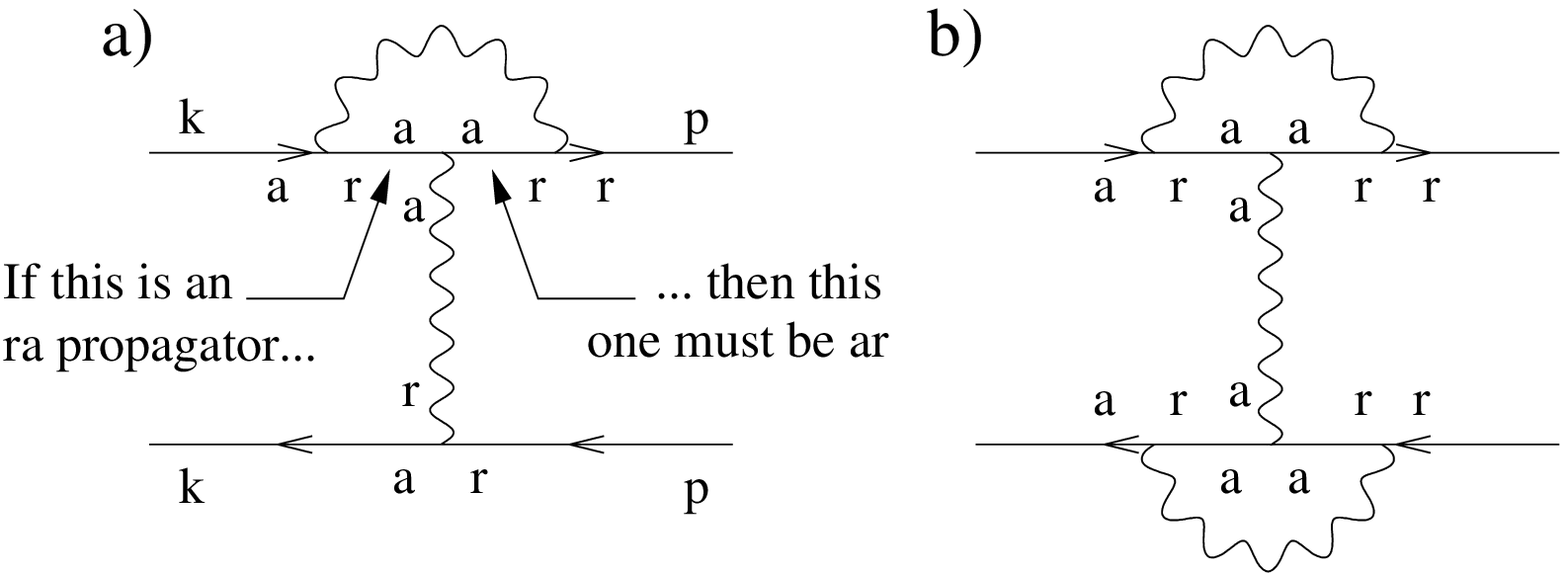}}
\caption{\label{fig:rungs_ra_basis_conductivity} Typical rung topologies for collinear processes expressed in the Keldysh basis.  To have a leading order contribution, the rungs must be ${\cal M}_{aarr}(k,-k,-p,p)$ or ${\cal M}_{rraa}(k,-k,-p,p)$ and must contain an alternation of $ra$ and $ar$ propagators.  Taking into account those constraints, it is easy to see that for all $r$,$a$ structures, there is always an $aa$ propagator in topology (b); it is thus suppressed compared to topology (a).}
\end{figure}

There is one further simplification that can be used to reduce the number of rungs in Fig.~\ref{fig:1to2_rungs_conductivity}.  To see this simplification, rungs must be expressed in the Keldysh basis.  In this basis, diagrams are expressed in terms of retarded, advanced and correlation functions (c.f. Eq.~(\ref{eq:physical_functions})) and it is thus easier to isolate pinching contributions.  This imposes constraints on the $r$,$a$ structure of rungs.  In particular, the 4-point rungs must be ${\cal M}(k,-k,-p,p) = {\cal M}_{aarr}(k,-k,-p,p)$ or ${\cal M}_{rraa}(k,-k,-p,p)$ in order for the side rails to pinch (a similar argument is used by the authors of \cite{Wang_Heinz_1999,Wang_Heinz_2003} to simplify their integral equation).  To have nearly pinching contributions from the collinear electron, we also need an alternation of retarded (ra) and advanced (ar) propagators on each side of the hard collinear photon.  Figure~\ref{fig:rungs_ra_basis_conductivity} shows typical rungs of Fig.~\ref{fig:1to2_rungs_conductivity} in the Keldysh basis submitted to the above constraints.  As can be seen from the figure, rungs with vertex corrections on both the upper and lower rails always contain an $aa$ propagator and are thus zero.  Other $r$,$a$ combinations are possible, but they do not pinch and are thus subleading.  In contrast, rungs with vertex corrections only on the upper or lower rail can have a pair of pinching propagators and are thus leading order.  

This infinite number of rungs with arbitrary number of vertex corrections must be resummed using another integral equation.  This is the diagrammatic implementation of the LPM effect \cite{Aurenche_etal_2000,AMY_2001,Baym_etal_2006}.  Physically, the LPM effect arises because of coherence effects in collinear photon emission/absorption.  In the process, the electron can suffer an arbitrary number of collisions with soft photons coming from the medium; since the formation time of the emitted/absorbed photon is of the same order as the mean free time between soft scatterings, then all the processes must be added coherently.  We will see more clearly in Sect.~\ref{sec:Integral_equations} how the rungs in Fig.~\ref{fig:1to2_rungs_conductivity} are related to these $1+N \rightarrow 2+N$ collinear processes.





\section{Derivation of the Integral Equation}
\label{sec:Integral_equations}

\begin{figure}
\resizebox{\textwidth}{!}{\includegraphics{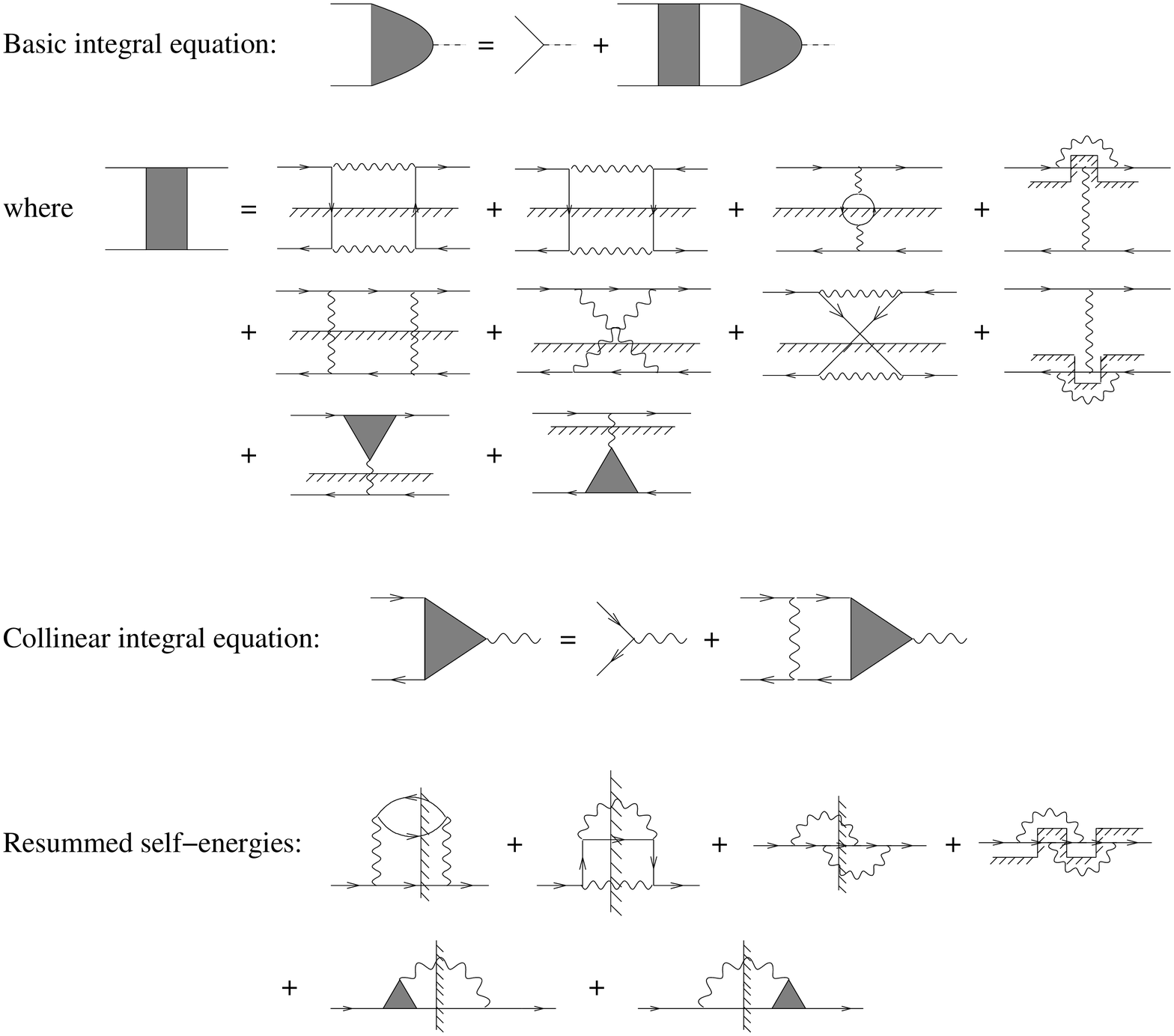}}
\caption{\label{fig:Summary} Diagrammatic summary of our leading order calculation of electrical conductivity in hot QED.  The basic integral equation (c.f. Eq.~(\ref{eq:integral_equation_schematic})) is due to the usual pinch singularities; its solution (represented by a grey half circle) must be substituted in the initial Kubo relation to get the conductivity.  All the rungs included in the kernel of the basic integral equation (represented by a grey rectangle) are shown.  These rungs can be divided in two categories: those corresponding to $2\rightarrow 2$ scatterings (first and second lines) and those corresponding to $1\rightarrow 2$ collinear scatterings (third line).  The collinear rungs represent an infinite number of vertex corrections (represented by a grey triangle) that are resummed using the collinear integral equation (c.f. Eq.~(\ref{eq:Vertex_resummation})).  Also shown are the self-energies that must be resummed in the side rail propagators in order to preserve gauge invariance, as dictated by Eq.~(\ref{eq:Constraint_M_charge}).}
\end{figure}

Sections \ref{sec:Derivation_constraint} and \ref{sec:Power_counting} show what ``rungs'', in the realm of all possible rungs, should be resummed in order to obtain the electrical conductivity at leading order.  A visual summary of the necessary rungs is shown in Fig.~\ref{fig:Summary}.  These rungs can be resummed using the integral equation~(\ref{eq:integral_equation_schematic}) in order to get the electrical conductivity.  As explained in Sect.~\ref{sec:Power_counting}, the rungs fall in two categories: those corresponding to $2\rightarrow 2$ scatterings (i.e. without collinear singularity) and those corresponding to $1\rightarrow 2$ collinear scatterings (i.e. containing collinear singularities).  Note that there is a finite/infinite number of rung types corresponding to $2\rightarrow 2$/$1\rightarrow 2$ scatterings, as stated in Sect.~\ref{sec:Complications}.

The goal now is to write down the appropriate integral equation and show its equivalence to the leading order results of Arnold, Moore and Yaffe \cite{AMY_2003a,AMY_2003b} obtained using an effective kinetic theory.  The plan is to separate the analysis of collinear and non-collinear rungs in equation~(\ref{eq:integral_equation_schematic}) according to ${\cal K} = ({\cal M}+{\cal N}){\cal F}$, where ${\cal N}$ and ${\cal M}$ correspond to the collinear and non-collinear rungs, respectively.

%

\subsection{Integral Equation Without Collinear Singularity (${\cal M}$)}
\label{sec:Integral_equation_without_collinear}

For simplicity, let us first consider the case without any collinear physics.  The diagrams that need to be resummed are shown in Fig. \ref{fig:Summary}.  To write down the integral equation, we follow the general diagrammatic method of Ref. \cite{Jeon_1995}.  In this method, all possible ways of cutting the effective vertex and the kernel are considered, resulting in a $4 \times 4$ integral matrix equation:
\begin{eqnarray}
\label{eq:4x4_integral_equation}
\mbox{\boldmath ${\cal D}$}_{F}^{\mu}(k) & = & \mbox{\boldmath ${\cal I}$}_{F}^{\mu}(k) + \int\frac{d^{4}p}{(2\pi)^{4}}\; \mbox{\boldmath ${\cal M}$}(k,p) \mbox{\boldmath ${\cal F}$}(p) \mbox{\boldmath ${\cal D}$}_{F}^{\mu}(p)
\end{eqnarray}
We use boldface letters here to emphasize the fact that $\mbox{\boldmath ${\cal D}$}$ and $\mbox{\boldmath ${\cal I}$}$ are 4 component column vectors and $\mbox{\boldmath ${\cal K}$} \equiv \mbox{\boldmath ${\cal M}{\cal F}$}$ is a  $4 \times 4$ matrix.  See Fig.~\ref{fig:4x4_integral_equation} for the graphical representation of this equation.  We again emphasize that, here and in the following, the order of the various fermionic components is not respected; for example, the effective vertex $\mbox{\boldmath ${\cal D}$}_{F}^{\mu}(p)$ has a Dirac structure and should be sandwiched between the two fermionic propagators contained in $\mbox{\boldmath ${\cal F}$}(p)$, something that is not apparent from the present notation.  Only explicit calculations show that the Dirac structure all works out.  The matrix $\mbox{\boldmath ${\cal M}$}$ corresponds to the ``rungs'' of the ladder diagrams and $\mbox{\boldmath ${\cal F}$}$ corresponds to the ``side rails''.  The explicit form of $\mbox{\boldmath ${\cal F}$}$ is:
\begin{figure}
\resizebox{\textwidth}{!}{\includegraphics{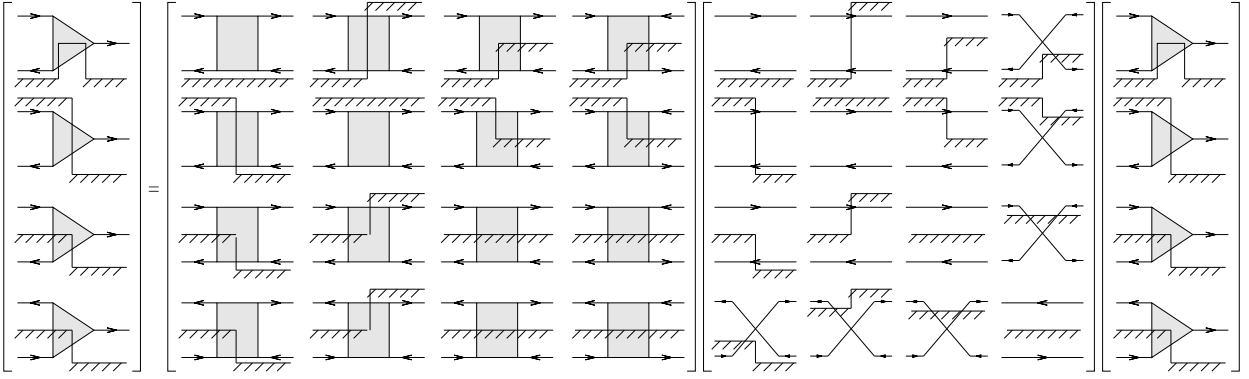}}
\caption{\label{fig:4x4_integral_equation} Graphical representation of the integral equation~(\ref{eq:4x4_integral_equation}).  The inhomogeneous term $\mbox{\boldmath ${\cal I}$}(p)$ is not shown explicitly.  It is a column vector with only the first component being nonzero, since it is an operator insertion and is thus point-like ({\it i.e.} it can only be in the shade or out of the shade).}
\end{figure}
\begin{eqnarray}
\label{eq:4x4_side_rail_matrix}
\mbox{\boldmath ${\cal F}$}(p) & = & \left( \begin{array}{cccc} G_{F}(p)G_{F}(p) & \Delta_{F}^{+}(p)\Delta_{F}^{-}(p) & G_{F}(p)\Delta_{F}^{-}(p) & \Delta_{F}^{+}(p)G_{F}(p) \\ \Delta_{F}^{-}(p)\Delta_{F}^{+}(p) & G_{F}^{*}(p)G_{F}^{*}(p) & \Delta_{F}^{-}(p)G_{F}^{*}(p) & G_{F}^{*}(p)\Delta_{F}^{+}(p) \\ G_{F}(p)\Delta_{F}^{+}(p) & \Delta_{F}^{+}(p)G_{F}^{*}(p) & G_{F}(p)G_{F}^{*}(p) & \Delta_{F}^{+}(p)\Delta_{F}^{+}(p) \\ \Delta_{F}^{-}(p)G_{F}(p) & G_{F}^{*}(p)\Delta_{F}^{-}(p) &\Delta_{F}^{-}(p)\Delta_{F}^{-}(p) & G_{F}^{*}(p)G_{F}(p) \end{array} \right) \nonumber \\
                               & = & \frac{1}{\Gamma_{p}}\left( \begin{array}{cccc} -2n\Delta_{F}^{+} & 2(1-n)\Delta_{F}^{-} & (1-2n)\Delta_{F}^{-} & (1-2n)\Delta_{F}^{+} \\ -2n\Delta_{F}^{+} & -2n\Delta_{F}^{+} & (1-2n)\Delta_{F}^{-} & (1-2n)\Delta_{F}^{+} \\ (1-2n)\Delta_{F}^{+} & (1-2n)\Delta_{F}^{+} & ((1-n)\Delta_{F}^{+}-n\Delta_{F}^{-}) & 2(1-n)\Delta_{F}^{+} \\ (1-2n)\Delta_{F}^{-} & (1-2n)\Delta_{F}^{-} & -2n\Delta_{F}^{-} & ((1-n)\Delta_{F}^{+}-n\Delta_{F}^{-}) \end{array} \right)
\end{eqnarray}
where we have used $n\equiv n_{F}(p^{0})$ and $\Delta_{F}^{\pm}\equiv \Delta_{F}^{\pm}(p^{0})$ and taken the pinch limit ($q\rightarrow 0$) in the last line.  In matrix form, Eq.~(\ref{eq:4x4_integral_equation}) has a complicated structure and is quite cumbersome.  In the scalar case, it is shown in \cite{Jeon_1995} that the $4\times 4$ integral matrix equation can be reduced to a one-dimensional integral equation using a special transformation.  This is expected, since one could start directly from the (one-dimensional) Euclidean integral equation, do the Matsubara sum and do the analytic continuation, resulting in a one-dimensional integral equation for real energies that should be equivalent to Eq.~(\ref{eq:4x4_integral_equation}) (see for example \cite{Basagoiti_2002} for such an approach).  Thus following Ref. \cite{Jeon_1995}, we decompose the side rail matrix~(\ref{eq:4x4_side_rail_matrix}) into outer products:
\begin{eqnarray}
\label{eq:Decomposition}
{\cal F}(p) & = & w(p)u^{T}(p) + h(p)j^{T}(p)
\end{eqnarray}
where the vectors $w(p)$, $u(p)$, $h(p)$ and $j(p)$ are given by:
\begin{eqnarray}
\label{eq:Outer_vectors}
w^{T}(p) & = & \frac{2n_{F}(p^{0})\Delta_{F}^{+}(p)}{\Gamma_{p}}\left( \begin{array}{cccc} 1 & 1 & \frac{(1-e^{\beta p^{0}})}{2} & \frac{(1-e^{-\beta p^{0}})}{2} \end{array} \right) \nonumber \\
u^{T}(p) & = & \left( \begin{array}{cccc} -1 & -1 & \frac{-(1-e^{-\beta p^{0}})}{2} & \frac{-(1-e^{\beta p^{0}})}{2} \end{array} \right) \nonumber \\
h^{T}(p) & = & \frac{2\rho_{F}^{+}(p)}{\Gamma_{p}} \left( \begin{array}{cccc} 0 & 0 & \frac{1}{4} & \frac{e^{-\beta p^{0}}}{4} \end{array} \right) \nonumber \\
j^{T}(p) & = & \left( \begin{array}{cccc} 0 & 0 & 1 & e^{\beta p^{0}} \end{array} \right)
\end{eqnarray}
This decomposition into outer products helps in simplifying Eq.~(\ref{eq:4x4_integral_equation}).  To see that, substitute Eq.~(\ref{eq:Decomposition}) into Eq.~(\ref{eq:4x4_integral_equation}) and iterate a few times.  Multiplying the result by $u^{T}(k)$ on both sides and noting that $j^{T}(p)\mbox{\boldmath ${\cal I}$}(p) = 0$, $u^{T}(p)h(p) = 0$, $j^{T}(p)w(p) = 0$, $j^{T}(p)\mbox{\boldmath ${\cal M}$}(p,l) \propto j^{T}(p)$ and $\mbox{\boldmath ${\cal M}$}(p,l)h(l) \propto h(p)$, it is possible to show that the term $h(p)j^{T}(p)$ in the decomposition~(\ref{eq:Decomposition}) gives no contribution when iterating.  The integral equation thus becomes:
\begin{eqnarray}
u^{T}(k)\mbox{\boldmath ${\cal D}$}_{F}^{\mu}(k) & = & u^{T}(k)\mbox{\boldmath ${\cal I}$}_{F}^{\mu}(k) + \int \frac{d^{4}p}{(2\pi)^{4}}  [u^{T}(k)\mbox{\boldmath ${\cal M}$}(k,p)w(p)][u^{T}(p)\mbox{\boldmath ${\cal D}$}_{F}^{\mu}(p)]
\end{eqnarray}
We can relabel the reduced effective vertex by ${\cal D}_{F}^{\mu}(k) = u^{T}(k)\mbox{\boldmath ${\cal D}$}_{F}^{\mu}(k)$.  Since only the first component of $\mbox{\boldmath ${\cal I}$}_{F}^{\mu}(k)$ is nonzero, we have $u^{T}(k)\mbox{\boldmath ${\cal I}$}_{F}^{\mu}(k) = -{\cal I}_{F}^{\mu}(k)$.  If we define ${\cal M}' = u^{T}(k)\mbox{\boldmath ${\cal M}$}(k,p)w(p)$ as our reduced kernel, the resulting integral equation becomes:
\begin{eqnarray}
\label{eq:Integral_equation_reduced_1}
{\cal D}_{F}^{\mu}(k) & = & -{\cal I}_{F}^{\mu}(k) + \int \frac{d^{4}p}{(2\pi)^{4}}{\cal M}'(k,p){\cal D}_{F}^{\mu}(p)
\end{eqnarray}
This last equation can be further reduced.  Using the condition of unitarity (\ref{eq:Unitarity}) to get rid of some ${\cal M}$ components, the KMS relations for 4-point functions (\ref{eq:KMS_4pt_function}) and the relations ${\cal M}_{02} = {\cal M}_{12}^{*}$, ${\cal M}_{20} = {\cal M}_{21}^{*}$, ${\cal M}_{22} = {\cal M}_{22}^{*}$ and ${\cal M}_{32} = {\cal M}_{32}^{*}$ (these last relations must be checked explicitly for each rung), a tedious calculation shows that Eq.~(\ref{eq:Integral_equation_reduced_1}) reduces to:
\begin{eqnarray}
\label{eq:Integral_equation_reduced_2}
{\cal D}_{F}^{i}(k) & = & -{\cal I}_{F}^{i}(k) + \int \frac{d^{4}p}{(2\pi)^{4}} \nonumber \\
            &   & \times \left[n_{F}(p^{0}) (e^{-\beta k^{0}}+1)\left({\cal M}_{22}(k,p)+e^{\beta k^{0}}{\cal M}_{32}(k,p)\right) (e^{\beta p^{0}}+1)\bar{\Delta}_{F}^{+}(p)\right] \frac{{\cal D}_{F}^{i}(p)}{\Gamma_{p}}
\end{eqnarray}
where the ${\cal M}_{ij}$'s ($i,j = 0,...,3$) refer to the matrix components of $\mbox{\boldmath ${\cal M}$}$ (correspond to the different ways of cutting the rung kernel, see Fig.~\ref{fig:4x4_integral_equation}), we defined $\Delta_{F}^{+}(p)\equiv p\!\!\!/\bar{\Delta}_{F}^{+}(p)$ and we used the fact that ${\cal D} \propto \gamma^{i}$.  Equation~(\ref{eq:Integral_equation_reduced_2}) is the fermionic version of Eq.~(\ref{eq:Integral_equation_Jeon}) first obtained in \cite{Jeon_1995} for scalar theories.  The fact that only ``completely cut'' components of ${\cal M}$ appear in Eq.~(\ref{eq:Integral_equation_reduced_2}) is consistent with the Ward identity constraint~(\ref{eq:Constraint_M_charge}).  However, it is still necessary to go through all the analysis to get the right proportionality factor, since the constraint~(\ref{eq:Constraint_M_charge}) is a proportionality relation when interpreted in terms of diagrams.  The reason why it is important to consider all possible cuts of ${\cal M}$ is because there are many ways to get an imaginary part at finite temperature; for example, we have $2\mbox{Im}\,\Sigma^{11} = \Sigma^{11}+\Sigma^{22} = \Sigma^{12} + \Sigma^{21}$ for any (time-ordered) two-point function (a consequence of unitarity).

As can be seen from Fig.~\ref{fig:2to2_rungs}, the rungs are all made of 4-point functions with two external vertices in the shade and two out of the shade.  Since cut propagators represent nearly on-shell thermal quasi-particles, the rungs can be naturally interpreted as $2\rightarrow 2$ scattering processes.  A tedious calculation shows that the sum of all the leading order rungs in Fig.~\ref{fig:2to2_rungs} can be converted into the square of a scattering matrix, where the scattering processes are given in Fig.~\ref{fig:2to2_scatterings}.  It is easy to see that diagrammatically, starting from the scattering processes (see caption of Fig.~\ref{fig:2to2_scatterings} for details).  In its present form, the effective vertex in Eq.~(\ref{eq:Integral_equation_reduced_2}) is a matrix in spinor space; for the Kubo relation (\ref{eq:Kubo_relation_schematic}) to be a scalar equation, we want ${\cal D}_{F}^{i}(k)$ to be a vector in spinor space.  Multiplying Eq.~(\ref{eq:Integral_equation_reduced_2}) from the left with $\bar{u}^{s}(\hat{k})$ (or equivalently $\bar{v}^{s}(\hat{k})$, see \cite{Basagoiti_2002} for their definition) and defining $D_{F}^{i}(k) \equiv \bar{u}^{s}(\hat{k}){\cal D}_{F}^{i}(k)$, we get:
\begin{eqnarray}
\label{eq:Integral_equation_reduced_3}
D_{F}^{i}(k) & = & -I_{F}^{i}(k) + \int \frac{d^{4}p}{(2\pi)^{4}} \frac{d^{4}l}{(2\pi)^{4}} \frac{d^{4}l'}{(2\pi)^{4}}\; (2\pi)^{4}\delta^{(4)}(k+p-l-l') (e^{-\beta k^{0}}+1) \nonumber \\
            &   & \hspace{0.7in} \times \left[\frac{1}{2}\sum_{v,m,n}^{f,s,h} |M_{uvmn}(k,p;l,l')|^{2}\; \bar{\Delta}_{m}^{+}(l)\bar{\Delta}_{n}^{+}(l')\bar{\Delta}_{F}^{+}(-p)\right] \frac{D_{F}^{i}(p)}{\Gamma_{p}}
\end{eqnarray}
where the $M_{uvmn}$'s are the $2\rightarrow 2$ scattering processes shown in Fig.~\ref{fig:2to2_scatterings}.  The sum is over flavors (f), species (s) and helicities (h).  Note that for electrical conductivity, $v$ can only be an electron or a positron, otherwise the resulting rung would not connect with the fermionic effective vertex (see Fig.~\ref{fig:integral_equation}).  Equation~(\ref{eq:Integral_equation_reduced_3}) is formally equivalent to the linearized Boltzmann equation obtained in \cite{AMY_2003a,AMY_2003b}, with all leading order $2\rightarrow 2$ scatterings included.  To show this more explicitly, we define the deviation from equilibrium $\chi^{i}(k) \equiv D^{i}(k)/\Gamma_k$ \cite{Jeon_1995,Jeon_Yaffe_1996} and substitute it into Eq.~(\ref{eq:Integral_equation_reduced_3}).  We also use the delta functions contained in the cut propagators to put the integrated momenta on-shell.  We remark here that at finite temperature, both positive and negative energies appear in the cut lines.  The resulting integral equation is:
\begin{eqnarray}
\label{eq:Integral_equation_reduced_4}
-I_{F}^{i}(k) & = & \Gamma_{k}\chi_{F}^{i}(k) + \int \frac{d^{3}p}{(2\pi)^{3}} \frac{d^{3}l}{(2\pi)^{3}} \frac{d^{3}l'}{(2\pi)^{3}}\; (2\pi)^{4}\delta^{(4)}(k+p-l-l')  \nonumber \\
            &   & \times \left[\frac{1}{2}\sum_{v,m,n}^{f,s,h} |M_{uvmn}(k,p;l,l')|^{2}\;\frac{n_{v}(p^{0})(1\pm n_{m}(l^{0}))(1\pm n_{n}({l'}^{0}))}{(1-n_{F}(k^{0}))}\; (\chi_{v}^{i}(p)-\chi_{m}^{i}(l)-\chi_{n}^{i}(l')) \right] \nonumber \\
\end{eqnarray}
with the understanding that $\chi_{\gamma} = 0$ (i.e. there is no bosonic effective vertex).  Using the fact that $\sum_{vmn}|M_{uvmn}(k,p;l,l')|^{2}$ is invariant under the change of labels $p \leftrightarrow -l'$, $l \leftrightarrow l'$ and the relation $\Gamma_{k} = (1+e^{-\beta k^{0}})\Sigma^{+}(k)$ (where $\Sigma^{+}(k)$ is a Wightman self-energy that can be expressed in terms of $\frac{1}{2}\sum_{vmn}|M_{uvmn}(k,p;l,l')|^{2}$), we finally get:
\begin{eqnarray}
\label{eq:Integral_equation_reduced_5}
-(1-n_{F}(k^{0}))I_{F}^{i}(k) & = & \frac{1}{2}\int \frac{d^{3}p}{(2\pi)^{3}} \frac{d^{3}l}{(2\pi)^{3}} \frac{d^{3}l'}{(2\pi)^{3}}\;  (2\pi)^{4}\delta^{(4)}(k+p-l-l')\; \sum_{v,m,n}^{f,s,h} |M_{uvmn}(k,p;l,l')|^{2} \nonumber \\
            &   & \times n_{F}(k^{0})n_{v}(p^{0})(1\pm n_{m}(l^{0}))(1\pm n_{n}({l'}^{0})) \nonumber \\
	    &   & \times \left[\chi^{i}(k) + \chi_{v}^{i}(p) - \chi_{m}^{i}(l) - \chi_{n}^{i}(l')\right]
\end{eqnarray}
This last equation is identical to the one obtained by Arnold, Moore and Yaffe \cite{AMY_2003a,AMY_2003b} using effective kinetic theory.

\subsection{Integral Equation With Collinear Singularities (${\cal N}$)}
\label{sec:Integral_equation_with_collinear}

\begin{figure}
\resizebox{5in}{!}{\includegraphics{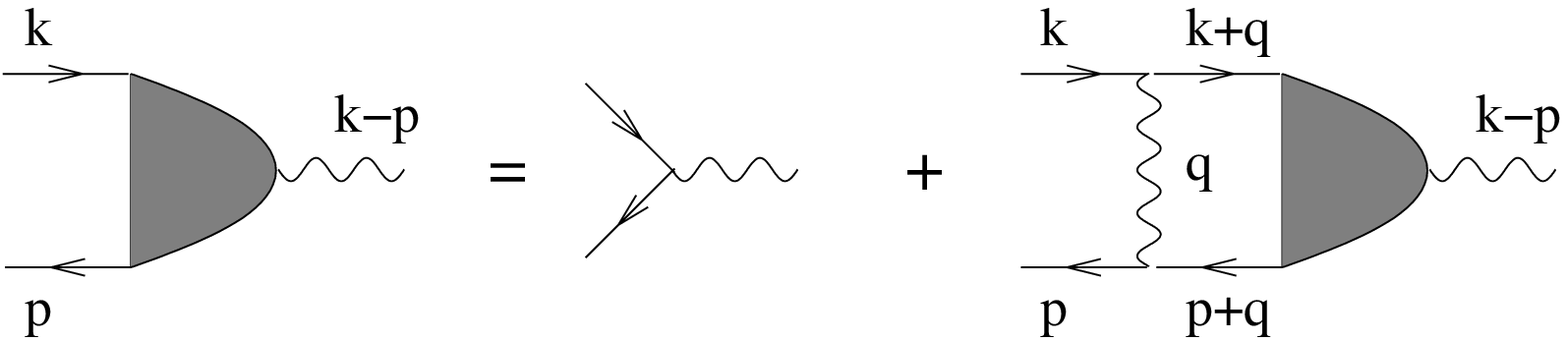}}
\caption{\label{fig:Integral_equation_collinear} Illustration of the integral equation~(\ref{eq:Vertex_resummation}).  The kernel ${
\cal N}_{\rm coll}$ is made of only a single soft photon exchange.}
\end{figure}

The case with collinear singularities can be studied with the tools developed in the previous section.  The only difference comes from the cut rungs that must be included in the $4\times 4$ integral equation.  In Fig.~\ref{fig:4x4_integral_equation}, the rung matrix is made of 4-point functions cut in all possible ways.  For collinear singularities, power counting shows that an infinite number of 3-point functions, corresponding to an infinite number of vertex corrections (see Fig.~\ref{fig:1to2_rungs_conductivity_2}), is needed at leading order.  Fortunately, this infinite number of 3-point functions has a simple structure and is thus manageable.  In analogy to the ``usual'' transport coefficient calculation, we can define an effective vertex ${\cal V}_{F}(k,p)$ that satisfies the following integral equation:
\begin{figure}
\resizebox{\textwidth}{!}{\includegraphics{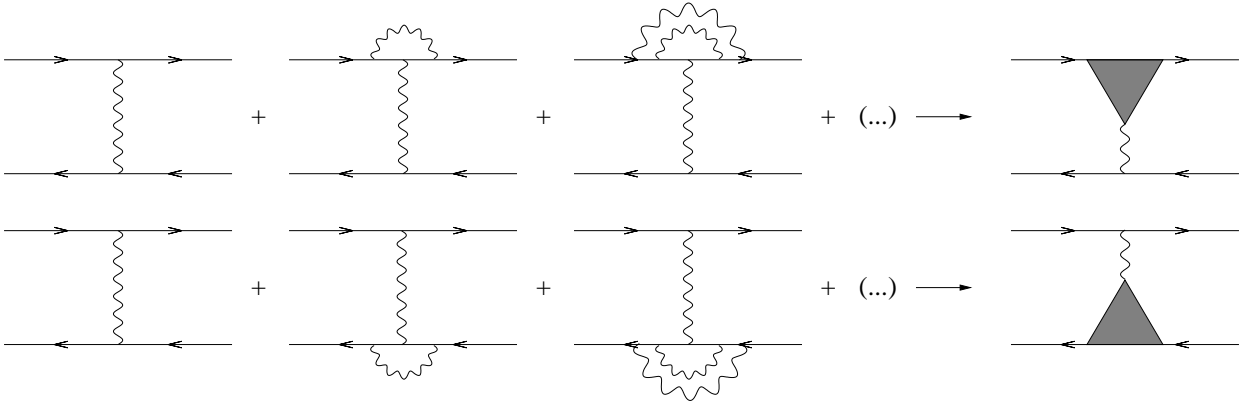}}
\caption{\label{fig:1to2_rungs_conductivity_2} Leading order rungs necessary for the calculation of electrical conductivity.  This infinite series of rungs can be resummed into an effective vertex ${\cal V}_{F}(k,p)$ using the integral equation~(\ref{eq:Vertex_resummation}).}
\end{figure}
\begin{eqnarray}
\label{eq:Vertex_resummation}
{\cal V}_{F}(k,p) & = & {\cal I}_{F}(k,p) + \int\frac{d^{4}q}{(2\pi)^{4}} {\cal N}_{\rm coll}(k,p,q){\cal F}(k,p,q){\cal V}_{F}(p,q)
\end{eqnarray}
where ${\cal N}_{\rm coll}$ is a rung with a single soft photon exchange and the external photon is collinear with the electron.  See Fig.~\ref{fig:Integral_equation_collinear} for an illustration of this integral equation.  This resummation is relevant for photon production including the LPM effect and is done in great detail in \cite{AMY_2001,AMY_2001b}.  Instead of using the closed-time-path or ``1-2'' formalism, they use the r/a formalism and are able to put the integral equation in a form convenient for numerical purposes.

Coming back to our initial problem, we need to find the components of the $4\times 4$ rung matrix. In the case of 4-point functions, the 16 entries of the rung matrix are given by the different ways of cutting a rung with 4 external vertices.  The collinear rungs considered in Fig.~\ref{fig:1to2_rungs_conductivity_2} are 3-point functions with one internal vertex.  The cuts coming from the three external vertices fill $2^{3} = 8$ entries of the rung matrix and each entry has two cuts due to the internal vertex.  Since the rung matrix is not completely filled, the reduction procedure used in Sect.~\ref{sec:Integral_equation_without_collinear} to get Eq.~(\ref{eq:Integral_equation_reduced_2}) could be spoiled and the whole technique would break down.  Fortunately this is not a problem, because each of the two topologies considered in Fig.~\ref{fig:1to2_rungs_conductivity_2} fill different parts of the rung matrix.  It is easy to check that the discussion preceding Eq.~(\ref{eq:Integral_equation_reduced_2}) is still valid in the present case.

\begin{figure}
\resizebox{3in}{!}{\includegraphics{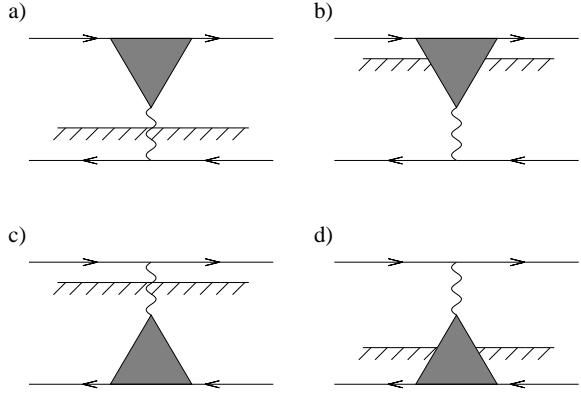}}
\caption{\label{fig:Collinear_cuts} Cut rungs left after the reduction procedure of Sect.~\ref{sec:Integral_equation_without_collinear}.  The cuts in the left column are included in ${\cal N}_{22}$; the others were already included in the non-collinear analysis.}
\end{figure}

Applying the reduction procedure of the preceding section to the topologies shown in Fig.~\ref{fig:1to2_rungs_conductivity_2}, we thus get an equation similar to Eq.~(\ref{eq:Integral_equation_reduced_2}):
\begin{eqnarray}
\label{eq:Integral_equation_collinear_reduced_2}
{\cal D}_{F}^{i}(k) & = & -{\cal I}_{F}^{i}(k) + \int \frac{d^{4}p}{(2\pi)^{4}} \left[n_{F}(p^{0}) (e^{-\beta k^{0}}+1){\cal N}_{22}(k,p) (e^{\beta p^{0}}+1)\bar{\Delta}_{F}^{+}(p)\right] \frac{{\cal D}_{F}^{i}(p)}{\Gamma_{p}}
\end{eqnarray}
where the ${\cal N}_{ij}$'s ($i,j = 0,...,3$) refer to the matrix components of $\mbox{\boldmath ${\cal N}$}$ (correspond to the different ways of cutting the rung kernel, see Fig.~\ref{fig:4x4_integral_equation}).  Note that ${\cal N}_{32}$ is automatically zero for a 3-point function; this is why it is not present in the above equation.  The only cuts left after the reduction procedure are shown in Fig.~\ref{fig:Collinear_cuts}.  Of the four cuts shown, only the ones in the left column are included in ${\cal N}_{22}$; the ones in the right column are part of the $2 \rightarrow 2$ scatterings considered in Sect.~\ref{sec:Integral_equation_without_collinear} and are left out of the collinear analysis.
The collinear rungs (a) and (c) can be written as ${\cal N}_{22\;(a)} = [-i{\cal V}_{F}^{\nu}]\Delta_{B\;\mu\nu}^{+}(k-p)[-iV_{1\rightarrow 2}^{\mu}]^{*}$ and ${\cal N}_{22\;(c)} = [-iV_{1\rightarrow 2}^{\mu}]\Delta_{B\;\mu\nu}^{+}(k-p)[-i{\cal V}_{F}^{\nu}]^{*}$, {\it i.e.} as a multiplication of a bare $1\rightarrow 2$ vertex $V_{1\rightarrow 2}^{\mu}$ and an effective vertex ${\cal V}_{F}^{\nu}$.  Note that rungs (a) and (c) are complex conjugate of each other; this is a consequence of the trace over the fermion structure in the Kubo formula, which allows us in this case to freely interchange the Dirac structure of the two vertices.  The sum of ${\cal N}_{22\;(a)}$ and ${\cal N}_{22\;(c)}$ is thus twice the real part of ${\cal N}_{22\;(a)}$.  Following the procedure of Sect.~\ref{sec:Integral_equation_without_collinear}, we multiply the left hand side of Eq.~\ref{eq:Integral_equation_collinear_reduced_2} with $\bar{u}^{s}(\hat{k})$ and get:
\begin{eqnarray}
\label{eq:Integral_equation_collinear_reduced_3}
D_{F}^{i}(k) & = & -I_{F}^{i}(k) + \int \frac{d^{4}p}{(2\pi)^{4}} \frac{d^{4}l}{(2\pi)^{4}}\; (2\pi)^{4}\delta^{(4)}(k-p-l) (e^{-\beta k^{0}}+1) \nonumber \\
            &   & \hspace{0.7in} \times \left[\sum_{v,m}^{f,s,h} 2\mbox{Re}[{\cal V}_{F}^{\mu}V_{1\rightarrow 2\;\mu}^{*}]_{vm}(k;p,l)\; \Delta_{B}^{+}(l)\bar{\Delta}_{F}^{+}(p)\right] \frac{D_{F}^{i}(p)}{\Gamma_{p}}
\end{eqnarray}
where $D_{F}^{i}(k) \equiv \bar{u}^{s}(\hat{k}){\cal D}_{F}^{i}(k)$.  The expression $[{\cal V}_{F}^{\mu}V_{1\rightarrow 2\;\mu}^{*}]$ can be viewed as $1\rightarrow 2$ matrix elements squared, where one of the element is an effective amplitude that takes into account the LPM effect.  This equation is formally equivalent to the linearized Boltzmann equation of Arnold, Moore and Yaffe \cite{AMY_2003a,AMY_2003b}, including the $1\rightarrow 2$ collinear scatterings and the LPM effect.  To make a closer connection to their results, we again define the deviation from equilibrium $\chi^{i}(k) \equiv D^{i}(k)/\Gamma_k$ and follow the procedure in Sect.~\ref{sec:Integral_equation_without_collinear}.  The final result is:
\begin{eqnarray}
\label{eq:Integral_equation_collinear_reduced_5}
-(1-n_{F}(k^{0}))I_{F}^{i}(k) & = & \frac{1}{2}\int \frac{d^{3}p}{(2\pi)^{3}} \frac{d^{3}l}{(2\pi)^{3}}\;  (2\pi)^{4}\delta^{(4)}(k-p-l)\; \sum_{v,m}^{f,s,h} 2\mbox{Re}[{\cal V}_{F}^{\mu}V_{1\rightarrow 2\;\mu}^{*}]_{vm}(k;p,l) \nonumber \\
           &   & \times \left[ n_{F}(k^{0})(1\pm n_{v}(p^{0}))(1\pm n_{m}(l^{0}))\left(\chi^{i}(k) - \chi_{v}^{i}(p) - \chi_{m}^{i}(l)\right)\right] \nonumber \\
	   &   & \int \frac{d^{3}p}{(2\pi)^{3}} \frac{d^{3}l}{(2\pi)^{3}}\;  (2\pi)^{4}\delta^{(4)}(k+p-l)\; \sum_{v,m}^{f,s,h} 2\mbox{Re}[{\cal V}_{F}^{\mu}V_{1\rightarrow 2\;\mu}^{*}]_{vm}(k,p;l) \nonumber \\
           &   & \times \left[ n_{F}(k^{0})n_{v}(p^{0})(1\pm n_{m}(l^{0}))\left(\chi^{i}(k) + \chi_{v}^{i}(p) - \chi_{m}^{i}(l)\right)\right]
\end{eqnarray}
This last equation is identical to the linearized Boltzann equation with the transverse momenta non-integrated of Arnold, Moore and Yaffe \cite{AMY_2003a,AMY_2003b} obtained using kinetic theory.

\section{Conclusion}
\label{sec:Conclusion}

In this paper, we  have derived the integral equations needed for the calculation of electrical conductivity in QED starting from basic quantum field theory.  A visual summary of our calculation is presented in Fig.~\ref{fig:Summary}.  With power counting arguments and a constraint on the ladder kernel due to the Ward identity, we have included all the necessary rungs for a leading order result.  Our calculation includes rungs corresponding to $2\rightarrow 2$ and $1\rightarrow 2$ collinear scatterings, including the LPM effect. An important point of our method is that since we used the Ward identity from the onset, gauge invariance is explicitly enforced throughout the calculation.  Specifically, the constraint on the ladder kernel tells us which rungs must be included in the integral equation for each self-energies resummed in the propagators.  We have finally shown the equivalence between our results and the ones of Arnold, Moore and Yaffe obtained from an effective kinetic theory.

As a final remark let us mention that, even if the present paper is restricted to the case of electrical conductivity in hot QED, the Ward identity constraint can in principle be applied to other gauge theories and other transport coefficients.  Of course, to generalize our calculation to non-abelian gauge theories is challenging, because of the presence of ghosts and of an even subtler power counting.  This is the subject of future research.  The shear viscosity case is also more complicated than the electrical conductivity case, due to the presence of both fermionic and bosonic effective vertices leading to coupled integral equations.  The extension of our calculation to shear viscosity is in progress \cite{Gagnon_Jeon_2006}.



\begin{acknowledgments}
The authors would like to thank Guy Moore for very illuminating discussions and his careful proofreading of the manuscript.  We also thank Gert Aarts, Jose M. Mart\'{i}nez Resco, Fran\c{c}ois Gelis, Simon Turbide and Fran\c{c}ois Fillion-Gourdeau for useful discussions.  This work was supported in part by the Natural Sciences and Engineering Research Council of Canada, and in part, for S.J., by le Fonds Nature et Technologies du Qu\'ebec.  S.J.~also thanks RIKEN BNL Center and U.S. Department of Energy [DE-AC02-98CH10886] for providing facilities essential for the completion of this work.
\end{acknowledgments}







\appendix

\section{Analytic Continuation of the Integral Equation}
\label{app:Analytic_continuation}

In this appendix we show in more detail the steps needed to go from the integral equation in Euclidean space~(\ref{eq:Integral_equation_Euclidean_modified}) to the one in Minkowski space~(\ref{eq:Integral_equation_Minkowski}).  As is well known from finite temperature field theory (see for example \cite{LeBellac_2000}), one must do the sum over Matsubara frequencies before doing the analytic continuation from imaginary energies to real energies.  An elegant way of doing this sum is presented in Ref. \cite{Basagoiti_2002}.  Briefly, the method gives a formula for the summation of expressions of the type
\begin{figure}
\resizebox{5in}{!}{\includegraphics{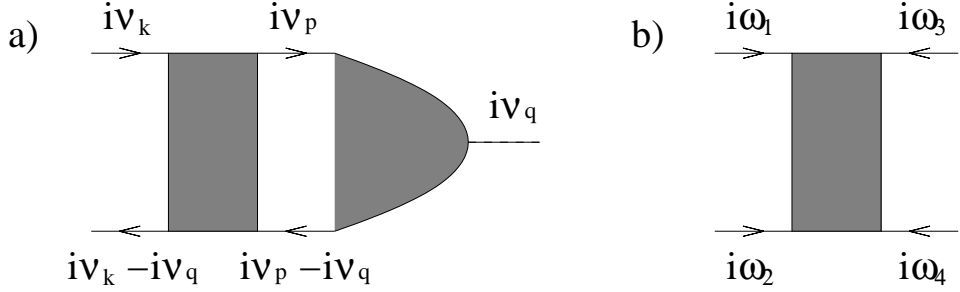}}
\caption{\label{fig:momentum_convention_integral_equation} Momentum convention used for a)~The summation formula~(\ref{eq:Summation_formula}) and the integral equation~(\ref{eq:Discrete_integral_equation}) b)~The general 4-point function~(\ref{eq:4_point_function_momentum})}.
\end{figure}
\begin{eqnarray}
T\sum_{i\nu_{p}} F(i\nu_{p}) & = & T\sum_{i\nu_{p}} {\cal M}(i\nu_{k},i\nu_{p},i\nu_{q})G_{F}(i\nu_{p}){\cal D}_{F}(i\nu_{p},i\nu_{p}-i\nu_{q})G_{F}(i\nu_{p}-i\nu_{q}) 
\end{eqnarray}
where $G_{F}(i\nu_{p})$ and $G_{F}(i\nu_{p}-i\nu_{q})$ are resummed propagators (the momentum convention is shown in Fig.~\ref{fig:momentum_convention_integral_equation}) and the $i\nu$'s are discrete frequencies.  This kind of expression is precisely what appears in the integral equation~(\ref{eq:Integral_equation_Euclidean_modified}), since the two propagators on the ``side rails'' are resummed (to regularize pinch singularities).  The idea is to replace the sum over discrete frequencies by a contour integration with a contour encircling those frequencies; the contour is then deformed to encompass the branch cuts coming from the resummed propagators and other possible poles contained in $F(z)$.  The resulting summation formula is \cite{Basagoiti_2002}
\begin{equation}
\label{eq:Summation_formula}
T\sum_{{\rm even/odd} \;\nu_{p}} F(i\nu_{p})  =  \mp\sum_{\rm poles} n_{B/F}(z_{i})\mbox{Res}(F,z = z_{i}) \pm\sum_{\rm cuts} \int_{-\infty}^{\infty}\frac{d\xi}{(2\pi i)} n_{B/F}(\xi)\mbox{Disc}(F)
\end{equation}
where $\xi$ is a parameter specifying the position along the branch cut.  

To be able to do the Matsubara sum present in Eq.~(\ref{eq:Integral_equation_Euclidean_modified}), we need to know the analytic structure of ${\cal D}_{F}$ and ${\cal M}$.  As argued in \cite{Basagoiti_2002} from induction, ${\cal D}_{F}$ has only the singularities of the product of $G_{F}(i\nu_{p})G_{F}(i\nu_{p}-i\nu_{q})$ and thus does not contribute to the sum.  On the other hand, the ${\cal M}$ factor contains poles and must be considered in the sum.  The problem is that we do not know a priori the precise form of the kernel ${\cal M}$, since this is the object we want to put a constraint on.  The way around this difficulty is to replace ${\cal M}$ with the spectral representation of a general 4-point function, of which we know the pole structure without specifying its exact form.  The program is thus to express the kernel ${\cal M}$ in spectral form, insert it in the integral equation ~(\ref{eq:Integral_equation_Euclidean_modified}), do the sum over Matsubara frequencies using Eq.~(\ref{eq:Summation_formula}) and do the analytic continuation towards real enegies.

In the imaginary-time formalism of finite temperature field theory, the (Euclidean) 4-point function is the thermal average of a time-ordered product of four bosonic/fermionic fields (in the following, we suppress spatial variables for simplicity):
\begin{eqnarray}
{\cal M}(\tau_{1},\tau_{2},\tau_{3},\tau_{4}) & = & \langle T(\phi_{a_{1}}(\tau_{1})\phi_{a_{2}}(\tau_{2})\phi_{a_{3}}(\tau_{3})\phi_{a_{4}}(\tau_{4}))\rangle
\end{eqnarray}
where the $\tau_{i}$'s are imaginary times and the $\phi_{a_{i}}$ are bosonic/fermionic fields.  The $a_{i}$'s are labels that denote the temporal order of the fields.  Since there are four fields, the $T$-product can be expanded into $4!$ possible time orderings.  For simplicity, let us consider a particular time ordering ($\tau_{1}>\tau_{4}>\tau_{2}>\tau_{3}$):
\begin{eqnarray}
{\cal M}_{a_{1}a_{4}a_{2}a_{3}} & = & \theta(\tau_{1}-\tau_{4})\theta(\tau_{4}-\tau_{2})\theta(\tau_{2}-\tau_{3})\langle \phi_{a_{1}}(\tau_{1})\phi_{a_{4}}(\tau_{4})\phi_{a_{2}}(\tau_{2})\phi_{a_{3}}(\tau_{3})\rangle 
\end{eqnarray}
Extracting the time dependence of the Heisenberg operators and averaging with respect to an equilibrium density matrix $\rho = e^{-\beta \hat{H}}$, we get
\begin{eqnarray}
{\cal M}_{a_{1}a_{4}a_{2}a_{3}} & = & \theta(\tau_{1}-\tau_{4})\theta(\tau_{4}-\tau_{2})\theta(\tau_{2}-\tau_{3}) \nonumber \\
                                                                     &   & \times \mbox{Tr}\left(e^{-\beta \hat{H}}e^{\hat{H}\tau_{1}}\phi_{a_{1}}(0)e^{\hat{H}(\tau_{4}-\tau_{1})}\phi_{a_{4}}(0)e^{\hat{H}(\tau_{2}-\tau_{4})}\phi_{a_{2}}(0)e^{\hat{H}(\tau_{3}-\tau_{2})}\phi_{a_{3}}(0)e^{-\hat{H}\tau_{3}}\right) \nonumber \\
\end{eqnarray}
Inserting complete sets of energy eigenstates (with eigenvalues $E_{m_{i}}$) between each bosonic/fermionic operator and rearranging, we obtain
\begin{eqnarray}
\label{eq:4_point_function_real}
{\cal M}_{a_{1}a_{4}a_{2}a_{3}} & = & \theta(\tau_{1}-\tau_{4})\theta(\tau_{4}-\tau_{2})\theta(\tau_{2}-\tau_{3}) \nonumber \\
                                &   & \times \sum_{m_{i}}\left(e^{-E_{m_{1}}(\beta+\tau_{3}-\tau_{1})}e^{-E_{m_{4}}(\tau_{1}-\tau_{4})}e^{-E_{m_{2}}(\tau_{4}-\tau_{2})}e^{-E_{m_{3}}(\tau_{2}-\tau_{3})}\right) \nonumber \\
                                &   & \times \langle m_{1}|\phi_{a_{1}}(0)|m_{4}\rangle \langle m_{4}|\phi_{a_{4}}(0)|m_{2}\rangle \langle m_{2}|\phi_{a_{2}}(0)|m_{3}\rangle \langle m_{3}|\phi_{a_{3}}(0)|m_{1}\rangle 
\end{eqnarray}
where $i$ runs from 1 to 4.  To get other time orderings, we only need to make permutations of the $a_{i}$'s and the $\tau_{i}$'s.  Since the integral equation~(\ref{eq:Integral_equation_Euclidean_modified}) is in momentum space, we need the spectral representation ${\cal M}$ in momentum space.  We thus take the Fourier transform of Eq.~(\ref{eq:4_point_function_real}):
\begin{eqnarray}
{\cal M}_{a_{1}a_{4}a_{2}a_{3}}(i\omega_{1},i\omega_{2},i\omega_{3},i\omega_{4}) & = & \int_{0}^{\beta}d\tau_{1}\int_{0}^{\beta}d\tau_{2}\int_{0}^{\beta}d\tau_{3}\int_{0}^{\beta}d\tau_{4} \;e^{i\omega_{1}\tau_{1}}e^{i\omega_{2}\tau_{2}}e^{i\omega_{3}\tau_{3}}e^{i\omega_{4}\tau_{4}} \nonumber \\
&   & \times {\cal M}_{a_{1}a_{4}a_{2}a_{3}}(\tau_{1},\tau_{2},\tau_{3},\tau_{4})
\end{eqnarray}
Note that the Fourier transform goes from 0 to $\beta$ and not $-\infty$ to $\infty$, since the presence of the medium breaks Lorentz invariance and imposes periodic (anti-periodic) boundary conditions on the imaginary time.  Consequently, the frequencies $i\omega_{i}$ are discrete and are even (odd) for bosons (fermions).  Doing the Fourier transform and manipulating the expression, we get
\begin{eqnarray}
\label{eq:4_point_function_momentum}
\lefteqn{{\cal M}_{a_{1}a_{4}a_{2}a_{3}}(i\omega_{1},i\omega_{2},i\omega_{3},i\omega_{4}) = } \nonumber \\
&   & \delta_{\omega_{1},\omega_{2},\omega_{3},\omega_{4}} \left(\prod_{j}^{4} \int\frac{dk_{j}}{(2\pi)} \;(2\pi)\delta(k_{1}+k_{2}+k_{3}+k_{4})\right) \times \nonumber \\
&   & \left[\frac{\gamma_{1423}}{[i\omega_{4}+i\omega_{2}+i\omega_{3}-(k_{4}+k_{2}+k_{3})][i\omega_{2}+i\omega_{3}-(k_{2}+k_{3})][i\omega_{3}-k_{3}]} \right. \nonumber \\
&   & + \frac{\gamma_{4231}}{[i\omega_{2}+i\omega_{3}+i\omega_{1}-(k_{2}+k_{3}+k_{1})][i\omega_{3}+i\omega_{1}-(k_{3}+k_{1})][i\omega_{1}-k_{1}]} \nonumber \\
&   & + \frac{\gamma_{2314}}{[i\omega_{3}+i\omega_{1}+i\omega_{4}-(k_{3}+k_{1}+k_{4})][i\omega_{1}+i\omega_{4}-(k_{1}+k_{4})][i\omega_{4}-k_{4}]} \nonumber \\
\label{Complete_4pt_function}
&   & \left. + \frac{\gamma_{3142}}{[i\omega_{1}+i\omega_{4}+i\omega_{2}-(k_{1}+k_{4}+k_{2})][i\omega_{4}+i\omega_{2}-(k_{4}+k_{2})][i\omega_{2}-k_{2}]} \right]
\end{eqnarray}
where $\delta_{\omega_{1},\omega_{2},\omega_{3},\omega_{4}}$ is a Kronecker delta and we have introduced the 4-point thermal Wightman functions \cite{Evans_1992}
\begin{equation}
\label{eq:Thermal_Wightman_functions}
\gamma_{1234}(k_{1},k_{2},k_{3},k_{4})  \equiv  \sum_{m_{i}} e^{E_{m_{1}}\beta} \left(\prod_{j=1}^{4}\langle m_{j}|\phi_{a_{j}}(0)|m_{j+1}\rangle\right) \left(\prod_{j=1}^{3}(2\pi)\delta(k_{j}-(E_{j+1}-E_{j}))\right)
\end{equation}
satisfying the (generalized) KMS condition
\begin{eqnarray}
\label{eq:Generalized_KMS}
\gamma_{1234}(k_{1},k_{2},k_{3},k_{4}) & = & (-1)e^{\beta k_{1}}\gamma_{2341}(k_{1},k_{2},k_{3},k_{4})
\end{eqnarray}
Equation~(\ref{eq:4_point_function_momentum}) is identical to the one obtained in \cite{Evans_1992} and gives the pole structure of one possible time ordering and its circular permutations.  As before, we obtain other time orderings by permutation.  Note that we could have expressed Eq.~(\ref{eq:4_point_function_momentum}) in terms of 4-point spectral densities ({\it i.e.} linear combinations of 4-point thermal Wightman functions) \cite{Evans_1992}, but we prefer to leave it in this form.

With the spectral form~(\ref{eq:4_point_function_momentum}) (and all the other time orderings), it is possible to do the Matsubara sum in the integral equation~(\ref{eq:Integral_equation_Euclidean_modified}).  The summation should be done for all the 24 time orderings of the 4-point function, but for simplicity we only show the summation of the $\gamma_{1423}$ term.  Inserting the first term of Eq.~(\ref{eq:4_point_function_momentum}) in Eq.~(\ref{eq:Integral_equation_Euclidean_modified}) (with the momentum convention shown in Fig.~\ref{fig:momentum_convention_integral_equation}), we get
\begin{eqnarray}
\label{eq:Discrete_integral_equation}
Q_{\mu}{\cal D}_{F}^{\mu}(i\nu_{k},i\nu_{k}-i\nu_{q})  & = & T\sum_{\nu_{p}}\int\frac{d^{3}p}{(2\pi)^{4}} \left(\prod_{j}^{4} \int\frac{dl_{j}}{(2\pi)} \;(2\pi)\delta(l_{1}+l_{2}+l_{3}+l_{4})\right) \gamma_{1423} \nonumber \\
                                                       &   & \times \left[\frac{G_{F}(i\nu_{p})\left[Q_{\mu}{\cal D}_{F}^{\mu}(i\nu_{p},i\nu_{p}-i\nu_{q})\right]G_{F}(i\nu_{p}-i\nu_{q})}{[-i\nu_{k}-l_{423}][i\nu_{p}+i\nu_{k}-i\nu_{q}+l_{23}][i\nu_{p}+l_{3}]}\right]
\end{eqnarray}
where we dropped the inhomogeneous term (which vanishes in the $q_{\mu}\rightarrow 0$ limit) and we used the shorthand notation $l_{ij} \equiv l_{i}+l_{j}$.  Note that the notation is symbolic and does not respect the Dirac structure of the equation.  Using the summation formula~(\ref{eq:Summation_formula}) and remembering that $i\nu_{k}$ is odd, $i\nu_{p}$ is odd and $i\nu_{q}$ is even in $\pi$, we obtain
\begin{eqnarray}
\label{eq:Summed_integral_equation}
\lefteqn{Q_{\mu}{\cal D}_{F}^{\mu}(i\nu_{k},i\nu_{k}-i\nu_{q}) = } \nonumber \\ 
                                                       &   & \int\frac{d^{3}p}{(2\pi)^{4}} \left(\prod_{j}^{4} \int\frac{dl_{j}}{(2\pi)} \;(2\pi)\delta(l_{1}+l_{2}+l_{3}+l_{4})\right) \gamma_{1423} \nonumber \\
                                                       &   & \times \left[\frac{n_{F}(-i\nu_{k}+i\nu_{q}-l_{23}) D(-i\nu_{k}+i\nu_{q}-l_{23},-i\nu_{k}-l_{23})}{[-i\nu_{k}-l_{423}][-i\nu_{k}+i\nu_{q}-l_{2}]}   \right.  + \frac{n_{F}(-l_{3}) D(-l_{3},-i\nu_{q}-l_{3})}{[-i\nu_{k}-l_{423}][i\nu_{k}-i\nu_{q}+l_{2}]} \nonumber \\
                            &   & - \int_{-\infty}^{\infty}\frac{d\xi}{(2\pi i)} n_{F}(\xi) \left( \frac{D(\xi+i\epsilon,\xi+i\epsilon-i\nu_{q})}{[-i\nu_{k}-l_{423}][\xi+i\epsilon+i\nu_{k}-i\nu_{q}+l_{23}][\xi+i\epsilon+l_{3}]} \right. \nonumber \\
                            &   & \hspace{1.3in} - \left. \frac{D(\xi-i\epsilon,\xi-i\epsilon-i\nu_{q})}{[-i\nu_{k}-l_{423}][\xi-i\epsilon+i\nu_{k}-i\nu_{q}+l_{23}][\xi-i\epsilon+l_{3}]} \right)  \nonumber \\
                            &   & - \int_{-\infty}^{\infty}\frac{d\xi}{(2\pi i)} n_{F}(\xi) \left( \frac{D(\xi+i\epsilon+i\nu_{q},\xi+i\epsilon)}{[-i\nu_{k}-l_{423}][\xi+i\epsilon+i\nu_{k}+l_{23}][\xi+i\epsilon+i\nu_{q}+l_{3}]} \right. \nonumber \\
                            &   & \hspace{1.3in} - \left.\left. \frac{D(\xi-i\epsilon+i\nu_{q},\xi-i\epsilon)}{[-i\nu_{k}-l_{423}][\xi-i\epsilon+i\nu_{k}+l_{23}][\xi-i\epsilon+i\nu_{q}+l_{3}]} \right)\right]
\end{eqnarray}
where we used the shorthand notation $D(a,b) \equiv G_{F}(a)[Q_{\mu}{\cal D}_{F}^{\mu}(a,b)]G_{F}(b)$.  In the above expression, the propagator $G_{F}(l_{3})$ has no $i\epsilon$ prescription and is thus undefined.  This problem can be cured by assigning a small imaginary part to the integrated momenta $l_{i}$, provided that it does not violate the delta function constraint.  This is allowed since the 4-point function is an analytic function and it will remain so no matter how the imaginary parts are assigned \cite{Jeon_1993}.  In the present case, two possible choices are $l_{3} \rightarrow l_{3}\pm 2i\epsilon$; the remaining momenta can be adjusted in many ways so that the delta function constraint is satisfied.  It is important to note here that this internal assignment of imaginary parts must always be ``less important'' than the $i\epsilon$'s coming from the analytic continuation of $i\nu_{k}$ and $i\nu_{q}$, because the latter encode the physics of transport coefficients.  Doing the analytic continuation $i\nu_{k} \rightarrow k^{0}+2i\epsilon$, $i\nu_{q} \rightarrow q^{0}+4i\epsilon$ and doing the integration, the integral equation for ${\cal M}_{1423}$ becomes:
\begin{eqnarray}
\lefteqn{\lim_{q\rightarrow 0} q_{\mu}{\cal D}_{F}^{\mu}(k^{0}+i\epsilon,k^{0}-q^{0}-i\epsilon) = } \nonumber \\
                                                       &   & \int\frac{d^{3}p}{(2\pi)^{4}} \left(\prod_{j}^{4} \int\frac{dl_{j}}{(2\pi)} \;(2\pi)\delta(l_{1}+l_{2}+l_{3}+l_{4})\right) \gamma_{1423} \nonumber \\
                                                       &   & \times \left[\frac{-(n_{B}(-l_{23}) + n_{F}(-k^{0}-l_{23}))D(-k^{0}-l_{23}+i\epsilon,-k^{0}-l_{23}-i\epsilon)}{[k^{0}+l_{423}+i\epsilon][k^{0}+l_{2}-i\epsilon]}\right]
\end{eqnarray}
where non-pinching pieces have been dropped and the $q\rightarrow 0$ limit has been taken.  Note that the result is independent of the $i\epsilon$ prescription for $l_{3}$.  It can be shown by explicit calculations that this independence on the assignment of imaginary parts on internal momenta holds for all time orderings.

We can do the same exercise with other time orderings.  Collecting all terms together and rearranging, the complete Matsubara summed integral equation is (spatial variables are still suppressed here and the $q\rightarrow 0$ is implicit)
\begin{eqnarray}
\label{eq:Integral_equation_Matsubara_summed}
q_{\mu}{\cal D}^{\mu}(k^{0}) & = & \int\frac{d^{3}p}{(2\pi)^{3}} \int\frac{dp^{0}}{(2\pi)}\; (1-n_{F}(p^{0})) G_{F}^{\rm ret}(p^{0})q_{\mu}{\cal D}^{\mu}(p^{0})G_{F}^{\rm adv}(p^{0}) \nonumber \\
                             &   & \times \left[ (1+e^{\beta k^{0}}) \int\frac{dl_{1}}{(2\pi)}\int\frac{dl_{2}}{(2\pi)} \left(\frac{\gamma_{2341}-\gamma_{2314}+\gamma_{3241}-\gamma_{3214}}{[k^{0}+l_{2}-i\epsilon][k^{0}-l_{1}+i\epsilon]}\right) \right. \nonumber \\
\label{Matsubara_integral_equation}
			     &   & \; \left. + (1+e^{-\beta k^{0}}) \int\frac{dl_{1}}{(2\pi)}\int\frac{dl_{2}}{(2\pi)} \left(\frac{\gamma_{1324}+\gamma_{1342}-\gamma_{3124}-\gamma_{3142}}{[k^{0}+l_{2}-i\epsilon][k^{0}-l_{1}+i\epsilon]}\right) \right]
\end{eqnarray}
The arguments of the $\gamma$'s are $\gamma_{a_{1}a_{2}a_{3}a_{4}}(l_{1},l_{2},p^{0}-l_{1}+k^{0})$ on the second line and $\gamma_{a_{1}a_{2}a_{3}a_{4}}(l_{1},l_{2},p^{0}-l_{2}-k^{0})$ on the third line.  To make sense of this equation, let us compare it the the integral equation obtained by Jeon \cite{Jeon_1995} (suitably generalized to fermions):
\begin{eqnarray}
\label{eq:Integral_equation_Jeon}
q_{\mu}{\cal D}^{\mu}(k^{0}) & = & \int\frac{d^{4}p}{(2\pi)^{4}}\; \left[n_{F}(p^{0})(1+e^{\beta k^{0}})\left(M_{33}+e^{-\beta k^{0}}M_{23}\right)(1+e^{-\beta p^{0}}) \frac{q_{\mu}{\cal D}^{\mu}(p^{0})}{\Gamma_{p}}\right] \nonumber \\
                             & = &\int\frac{d^{4}p}{(2\pi)^{4}}\; (1-n_{F}(p^{0}))G_{F}^{\rm ret}(p)q_{\mu}{\cal D}^{\mu}(p)G_{F}^{\rm adv}(p) \nonumber \\
			     &   & \times \left[ (1+e^{\beta k^{0}}) n_{F}(p^{0})(1+e^{-\beta p^{0}})M_{33} \right. \nonumber \\ 
                             &   & \; \left. + (1+e^{-\beta k^{0}}) n_{F}(p^{0})(1+e^{-\beta p^{0}})M_{23} \right]
\end{eqnarray}
where $M_{33}$ and $M_{23}$ represent (completely cut) kernels.  Comparing Eqs.~(\ref{eq:Integral_equation_Matsubara_summed}) and (\ref{eq:Integral_equation_Jeon}), we clearly see that they have the same structure.  This is normal, since they both represent the same physics.  We can thus infer from this that the double integrals over $l_{1}$ and $l_{2}$ in Eq.~(\ref{eq:Integral_equation_Matsubara_summed}) are the general spectral representation of (completely cut) kernels.  The fact that kernels (4-point functions) are represented by double integrals (3-point functions) in Eq.~(\ref{eq:Integral_equation_Matsubara_summed}) is normal, because there are only two independent momenta in the $q\rightarrow 0$ limit.  The similarity between Eqs.~(\ref{eq:Integral_equation_Matsubara_summed}) and (\ref{eq:Integral_equation_Jeon}) also shows that the Euclidean integral equation keeps its form when going to Minkowski space.

\end{document}